\documentclass[twoside,twocolumn,english]{revtex4-2}
\usepackage[T1]{fontenc}
\usepackage[latin9]{inputenc}
\usepackage{color}
\usepackage{float}
\usepackage{dsfont}
\usepackage{amsmath}
\usepackage{amssymb}
\usepackage{graphicx}
\usepackage{babel}
\begin{document}
\title{Coexistence of 1D and 2D topology and genesis of Dirac cones in the
chiral Aubry-André model}
\author{T. V. C. Antão$^{1,2}$, Daniel Miranda$^{3}$ and N. M. R. Peres$^{3,4,5}$}
\address{$^{1}$Department of Applied Physics, Aalto University, 02150 Espoo,
Finland}
\address{$^{2}$Laboratorio de Instrumentação e Física Experimental de Partículas,
University of Minho, 4710-057 Braga, Portugal}
\address{$^{3}$Centro de Física das Universidade do Minho e do Porto (CFUMUP)
e Departamento de Física, Universidade do Minho, P-4710-057 Braga,
Portugal}
\address{$^{4}$International Iberian Nanotechnology Laboratory (INL), Av Mestre
José Veiga, 4715-330 Braga, Portugal}
\address{$^{5}$POLIMA - Center for Polariton-driven Light-Matter Interactions,
University of Southern Denmark, Campusvej 55, DK-5230 Odense M, Denmark}
\begin{abstract}
We construct a one-dimensional (1D) topological SSH-like model with
chiral symmetry and a superimposed hopping modulation, which we call
the chiral Aubry-André model. We show that its topological properties
can be described in terms of a pair $\left(C,W\right)$ of a two-dimensional
(2D) Chern number $C$, stemming from a superspace description of
the model, and a 1D winding number $W$, originating in its chiral
symmetric nature. Thus, we showcase for the first time explicit coexistence
of 1D and 2D topology in a model existing in 1D physical space. We
detail the superspace description by showcasing how our model can
be mapped to a Harper-Hofstadter model, familiar from the description
of the integer quantum Hall effect, and analyze the vanishing field
limit analytically. An extension of the method used for vanishing
fields is provided in order to handle any finite fields, corresponding
to hopping modulations both commensurate and incommensurate with the
lattice. In addition, this formalism allows us to obtain certain features
of the 2D superspace model, such as its number of massless Dirac nodes,
purely in terms of topological quantities, computed without the need
to go into momentum space.
\end{abstract}
\maketitle

\section{Introduction}

Quasicrystals, once the subject of controversy, have now been firmly
established as intriguing materials that defy the classification of
traditional crystals. Indeed, their mathematical description was proposed
a long time before their actual observation in 1982 and subsequent
publication in 1984 \citep{Shechtman_2013}. Since then, many forms
and realizations of quasicrystals have been reported \citep{Macia_2006}.
One reason for interest is that they evade the classification of usual
crystals based on symmetries of their underlying discrete translation
groups, but nonetheless are ``ordered'' in a translational sense
\citep{Steinhardt_1987,Senechal_1995}. In the present day, it has
long been realized that quasicrystals in $d$-dimensional space can
be thought of as crystalline structures living in a higher $d+d'$
dimensional superspace, which are projected down to a physical dimension,
and this feature has been utilized to explore the properties of topological
quasicrystals.

The method used for the projection can be picked from a variety of
approaches, the most common of which may be the so-called ``cut and
project'' method \citep{Duneau_null,Elser_1985} which is at the
heart of the most famous quasicrystalline structures, such as the
Fibonacci quasicrystal, which can be realized as a projection from
a 2D superspace lattice into a 1D chain \citep{Jagannathan_2021},
or the Penrose tiling, which is a two-dimensional slice of a five-dimensional
hypercubic superspace lattice \citep{Bruijn_1981}. Different approaches
include a ``twisting'' approach, based on coupling and twisting
of monolayers of certain materials into twisted van der Waals heterostructures,
such as twisted-bilayer graphene, \citep{Moon_2019}, for which incommensurability
has been studied and is argued to be an important feature \citep{Amorim2023,Crosse2021}.

In some approaches, one actually starts with a crystalline structure
in the physical space, upon which some periodic potential incommensurate
with the lattice is added, allowing the unfolding of the model into
a higher superspace description. This ``incommensurate potential''
method is at the heart of the so-called Aubry-André (AA) models \citep{Wu_2021,Dominguez-Castro_2018}. 

The fact that quasicrystalline structures, such as AA models, can
host non-trivial single-particle as well as many-body topology is
one of their most captivating aspects, from both technological and
theoretical perspectives. On the theory side, it has been pointed
out that the existence of phasonic excitations or degrees of freedom
corresponding, for instance, to momenta along synthetic dimensions
in superspace can be used to physically realize $d+d'$-dimensional
topological invariants in $d$-dimensional systems \citep{Kraus_2012,Kraus_2013,Lohse_2018,Prodan_2015}. 

The standard example of this feature is the presence of the two-dimensional
(2D) Chern number in a one-dimensional (1D) AA model. Much like the
``cut and project'' approach, where deformations of the cutting
window induce additional mechanical degrees of freedom, shifting an
AA potential relative to the lattice also corresponds to a so-called
phasonic degree of freedom, and a 2D superspace crystal can be built
by interpreting this phason as a momentum coordinate. This occurs
because a periodic potential of this kind can be though of as unfolding
the model into a series of replicas and generating hoppings between
them. The resulting lattice exists in one dimension higher, and the
additional dimension is said to be ``synthetic''. The periodicity
of the potential is mapped to the magnitude of a ``magnetic field''
applied to the higher dimensional lattice, which induces Peierls phases
when hopping along the synthetic dimension. Thus, the superspace model
of the AA model maps to a square lattice hosting integer quantum hall
effect (IQHE) physics \citep{Hofstadter_1976,Klitzing_1980}, where
transverse conductance is well known to be topologically characterized
by the Chern number \citep{Hatsugai_1993,Thouless_82}. Beyond these
simple models, the superspace picture and phasonic degrees of freedom
have also been used to characterize the many-body topology of quasicrystals
\citep{Else_2021}, where it is the topological terms which can appear
in a Lagrangian describing the elastic and phasonic deformations of
the superspace lattice that determine the possible topological properties
of the model.

The remarkable feature of quasicrystals we deal with here, is that
any physical quasicrystal model retains some of the topological properties
of the superspace model. This can be readily verified by analyzing
the spectrum and eigenstates of the AA model, which can be seen to
clearly exhibit finite energy states localized at the edges. Furthermore,
these are a direct consequence of the topological properties of the
presence of a Chern invariant. This crucial point has been studied
experimentally in the context of photonic crystals using waveguide
arrays within dielectric media \citep{Kraus_2012}.

Besides quasicrystals, chiral symmetric models have also been shown
to host extremely rich topology, and a propensity for generalization
into different topological models via a process of stacking. Chern
numbers characteristic of 2D matter can, for instance, exist in certain
quasi-1D models subject to magnetic fields, such as the Creutz-Su-Schrieffer-Heeger
(CSSH) models \citep{Zurita2021}, constructed by stacking two SSH
chains in a square configuration, and connecting them with diagonal,
second nearest neighbor, as well as vertical, nearest neighbor bonds.

Additionally, stacking multiples of these SSH chains, repeating the
Creutz-ladder pattern in the 2D plane or 3D space, one can construct
Weyl-semimetals in 2D and 3D \citep{Ganeshan2015}, and when certain
types of interactions are turned on, these systems have even been
shown to host topological superconducting properties \citep{Rosenberg2022}.
In such previous studies, AA-type models are considered, however in
order to analyze the spectra and other properties of these systems,
modulations have always been considered to be commensurate with the
lattice.

It has furthermore been shown that incommensurate disorder can drive an SSH model from a trivial into an Anderson topological insulator phase \citep{Longhi2020}.

We argue here that the idea of stacking topological models of low
dimension into a higher dimensional model is not dissimilar from the
aforementioned superspace description of quasicrystals, and thus the
presence of 2D Chern numbers in quasi-1D systems, the rich landscape
of SSH-type chiral symmetric models are suggestive of a question:
Is it possible to construct a topological crystalline existing physically
in 1D, and via modulating using an AA approach, render it doubly topological?
Or in other words, can the AA modulation of a chiral symmetric model
lead to the coexistence of 1D and 2D topology in a model in 1D physical
space?

We show in this work that topological invariants characteristic to
different dimensionalities can coexist in a physically meaningful
and experimentally testable fashion in a model we call the chiral Aubry-André model (cAA), extending previous work on this subject into the fully incommensurate setting. The cAA is perhaps one of the simplest conceivable candidates for such a coexistence of 1D and 2D topology, as it is built from the SSH model \citep{Su_1979} which exhibits topological properties protected by chiral symmetry. In particular, the SSH model hosts a quantized topological invariant corresponding to the winding number $W$.

\begin{figure}
\begin{centering}
\includegraphics[scale=0.45]{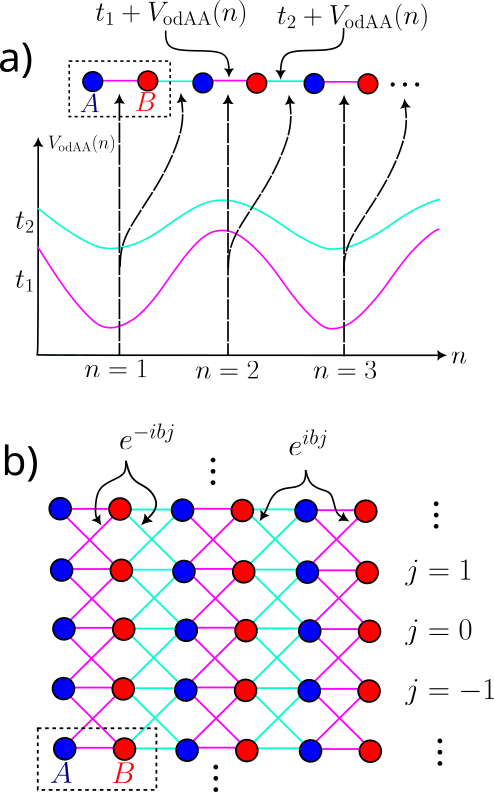}
\par\end{centering}
\caption{(a) Schematic of the cAA model accompanied by an example plot of the
co-sinusoidal hopping amplitude modulation of the underlying SSH model
$V_{\text{odAA}}(n)$ as a function of the lattice site. (b) Illustration
of the corresponding model using a synthetic dimension, capturing
all possible phasonic shifts of the Aubry-André potential. Blue and
red circles in either panel correspond to the A and B sublattice sites
respectively, horizontal lines connecting the sites indicate the physical
hoppings, and diagonal lines connecting adjacent copies of the model
are generated by the presence of the incommensurate hopping modulation.
Hoppings connecting replicas of the 1D chain are accompanied by Peierls
phases $e^{\pm ibj}$, where the inverse modulation periodicity $b$
plays the role of an applied magnetic field. \label{fig:Illustration} }
\end{figure}

In order to build the cAA, to the SSH model, an off-diagonal
Aubry-André (odAA) term is added. This results in a quasi-periodic
modulation of the hopping amplitudes. Global shifts in this modulation
correspond to phasonic degrees of freedom, providing the ground for
the topological enrichment of the model. A schematic illustration
of the cAA model is presented in Fig. \ref{fig:Illustration}.

\begin{figure}
\begin{centering}
\includegraphics[width=0.45\textwidth]{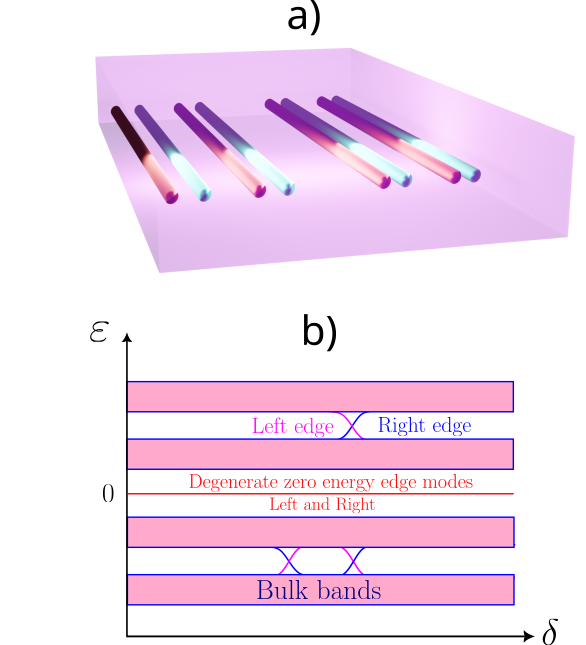}
\par\end{centering}
\caption{(a) Artistic rendition of the realization of the cAA model in a photonic
crystal platform. The physical spacing between adjacent waveguides
is tuned to take into account both the SSH and odAA modulations. (b)
Typical example plot for the spectrum of the superspace cAA model
with an incommensurate potential. The presence of doubly degenerate
zero energy edge modes along with edge states at finite energies points
to the main results of this work: the coexistence of 1D and 2D topology
in a purely 1D model. \label{fig:(a)-Schematic-of}}
\end{figure}

Experimental realization of our model using platforms such as photonic
crystals should not be out of reach, since the actual methodology
as utilized, by Kraus \textit{et. al.} \citep{Kraus_2012} (see Fig.
\ref{fig:(a)-Schematic-of}), for instance, could easily be adapted
to account for the different structure of hopping modulation necessary
to implement the cAA system. Many alternative platforms exist which
may realize the cAA model, such as twisted bilayer graphene with adhered
hydrogen or deuterium atoms. This has been proposed in the past as
a way to generate and tune the modulation of a $S=1/2$ Aubry-André-Heisenberg
chain \citep{Lado2019}.

In the simplest cases where the unperturbed model has no special symmetry,
the odAA model has been shown to be topologically equivalent to the
AA model by using a superspace prescription. Its topology is entirely
characterized by the same Chern number $C$, in the same manner as
the traditional AA model, and indeed, a topological equivalence to
the Fibonacci quasicrystal has also been established \citep{Kraus_2012_1}. 

In our cAA model, however, the presence of chiral symmetry in the
original model actually breaks the topological equivalence between
AA and odAA models, since an on-site disorder term would break this
chiral symmetry, and therefore a change is observed in the Altland-Zirnbauer
symmetry class of the 1D model.

As showcased in the typical example of the dispersion
of the cAA model in Fig. \ref{fig:(a)-Schematic-of} (b), a series
of puzzling characteristics appear, such as zero energy edge modes,
and finite energy edge states connecting bulk bands.

The aim of this paper is mainly to provide a unified
topological treatment of these features, relating them back to a pair
of topological invariants $(C,W)$, allowing for an explanation of
the aforementioned topological equivalence by noting that by breaking
chiral symmetry $W$ becomes an ill-defined quantity.

Furthermore, the relative simplicity of the cAA model results in the
possibility of obtaining analytical expressions for the spectrum in
the limit of vanishing field (large periodicity for the AA modulation).
By constructing the superspace model as showcased in Fig. \ref{fig:Illustration}
(b), and working within this limit, we show that an additional series
of interesting features appear, such as the genesis of a variable
and tunable number of anisotropic massless Dirac cones. The Dirac
cones can be fused together by changing the model's relative hopping
and modulation magnitudes, allowing for phase transitions between
topological semimetal, topological insulator and trivial insulator
phases. The phenomenology and physics of Dirac cones
is interesting in its own right, being responsible for the spur of
research in many quantum materials, such as graphene. This carbon based 2D material naturally
hosts Dirac Fermions \citep{Neto_2009}. Furthermore, in graphene, it is well-known that the presence of
two Dirac cones leads to a quantization of conduction at half-integer
multiples of $e^{2}/h$ (where $e$ is the electron charge and $h$
is Planck's constant) when an external magnetic field is applied.
This effect is often called the half-integer quantum hall effect (HIQHE).

On the other hand, in the absence of Dirac cones, a magnetic field induces
a quantization of conductance at integer multiples of $e^{2}/h$,
resulting in the integer quantum hall effect (IQHE) instead. 

In the cAA model when considering hopping modulations of large yet finite
wavelengths, the genesis of Dirac cones in the superspace cAA model,
together with the ability to induce semimetal to insulator or nodal
line semimetal phase transitions, results in the possibility of simulating
both the HIQHE and IQHE.

Finally, changing properties such as phasonic degrees
of freedom is found to be an extremely useful knob for the study and
applications of topologically protected edge modes. In particular,
we show that shifting the phason allows for turning these modes on
or off, both at zero and finite energy in the 1D physical realization
of the model, as well as change at which physical edge of the system
they appear in.

The remainder of this work is organized as follows: In section II
we give some details on our definition and construction of the chiral
Aubry-André model, in section III we analyze the topological 1D and
properties of the energy spectrum from the point of view of the superspace
model in the limit of vanishing field (long wavelength odAA modulations).
We also discuss a way to calculate the number of massless Dirac cones
by counting changes in the winding number as the momentum in the synthetic
direction is traversed. In section IV we extend our discussion to
the presence of finite fields, and using topological markers \citep{KITAEV_2006_2,Bianco_Resta_2011,Gersdorff_2021,Chen_2022,Sykes_2022},
explicitly showcase the coexistence of topological invariants in the
incommensurate limit, extending our computations of the number of
Dirac cones to the case when the periodicity of the odAA term explicitly
breaks the translational symmetry of the model.

\section{The Chiral Aubry-André model\label{sec:The-Chiral-Aubry-Andr=0000E9}}

The main aim of this work is to showcase a toy model which is topologically
non-trivial in character from the outset, but which under the presence
of an AA-type incommensurate modulation exhibits enriched topological
properties, stemming from its superspace description. In this section,
we motivate and build such a model. Let us consider what is arguably
the simplest 1D topologically non-trivial model: the SSH model. This
model is in the BDI class of the Altland-Zirnbauer classification
of topological invariants \citep{Altland_Zirnbauer_1997,Ryu_2010},
wherein a $\mathbb{Z}$-valued topological invariant is well defined,
and often referred to as the winding number. The quantization of this
winding number is assured by the presence of two distinct sites per
unit cell, of type $A$ and $B$, or equivalently by the existence
of two sublattices, together with hopping terms which connect only
$A$-type sites to $B$-type sites and vice-versa. This latter condition
can, in fact, be thought of as the definition of chiral symmetry (which
for this reason sometimes referred to as sublattice symmetry). For
simplicity, we shall assume that the fundamental length scale, corresponding
to the lattice spacing is $a=1$ in natural units, such that the SSH
Hamiltonian reads

\begin{equation}
H_{\text{SSH}}=-\sum_{n}t_{1}a_{n}^{\dagger}b_{n}+t_{2}b_{n}^{\dagger}a_{n+1}+\text{h.c.},\label{eq:SSH}
\end{equation}
where $a_{n}^{\dagger}(a_{n})$ create (annihilate) excitations in
the $A-$sublattice within the unit cell $n$, and equivalently for
$b_{n}^{\dagger}(b_{n})$ in the $B$ sublattice. $t_{1}$ represents
the intra-cell hopping strength, and $t_{2}$ the inter-cell hopping.
The topological properties of the model crucially rely on the presence
of chiral symmetry, and quantization of the winding number is robust
to perturbations only if the added perturbations
themselves respect this symmetry. If we wish to
design a model with an additional topological invariant,
we must therefore preserve chiral symmetry. To achieve this, an off-diagonal
Aubry-André (odAA) hopping modulation is considered, given by the
expression

\begin{align}
H_{\text{odAA}}(\delta)= & -\Delta_{1}\sum_{n}\cos\left(2\pi bn-\delta\right)a_{n}^{\dagger}b_{n}+\text{h.c.}\nonumber \\
 & -\Delta_{2}\sum_{n}\cos\left(2\pi bn-\delta\right)b_{n}^{\dagger}a_{n+1}+\text{h.c.},\label{eq:odAA}
\end{align}
where the parameters $\Delta_{1}$ and $\Delta_{2}$ denote the strengths
of the modulation in intra-cell and inter-cell hoppings, respectively.
The modulation occurs periodically with a period $L=1/b$. As we shall
later see, \textbf{$b$} plays the role of magnetic flux in a 2D tight-binding
model subject to an external magnetic field (the superspace model).
Both intra-cell and inter-cell hoppings are varied by the cosine term
equivalently within each unit cell. In other words, within each unit
cell, the hopping modulation occurs only due to the difference in
the strength parameters $\Delta_{1}$ and $\Delta_{2}$. The parameter
$\delta$ represents an overall shift in the hopping modulation, constituting
a phasonic degree of freedom. We define the chiral Aubry-André (cAA)
model as $H_{\text{CAA}}(\delta)\equiv H_{\text{SSH}}+H_{\text{odAA}}(\delta)$.

\begin{widetext}

\begin{figure}[H]
\begin{centering}
\includegraphics[width=0.7\textwidth]{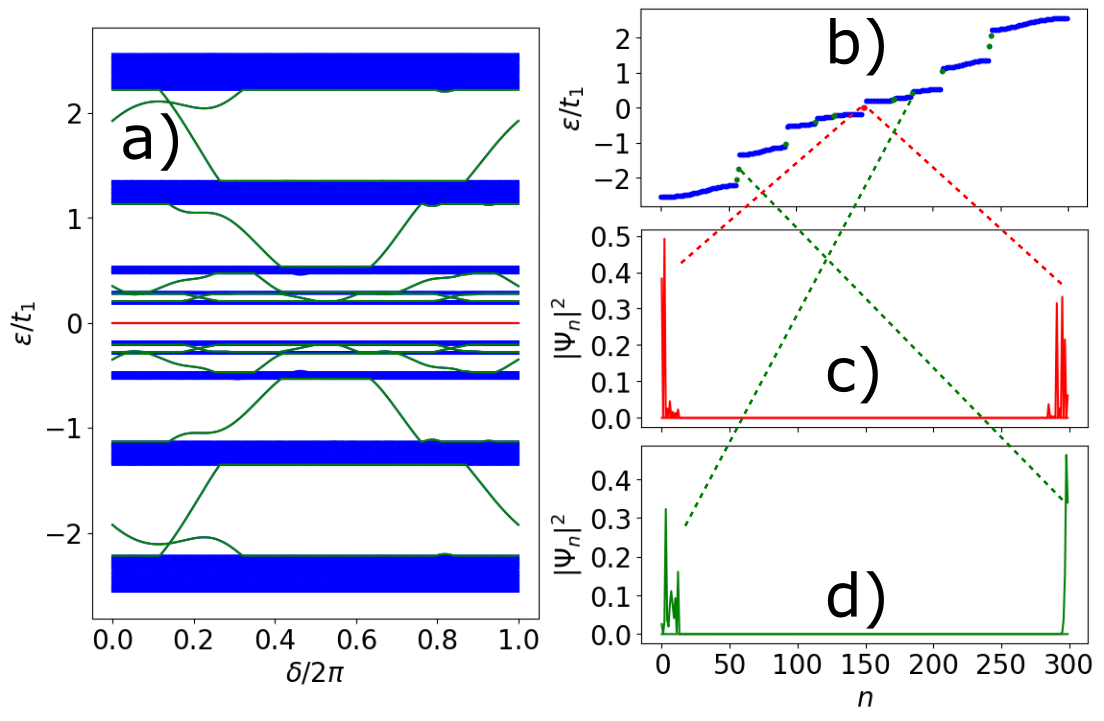}
\par\end{centering}
\caption{(a) Spectra of the cAA model as a function of the phasonic degree
of freedom or modulation shift $\delta$ with parameters $t_{1}=t_{2}=\Delta_{1}=1$
and $\Delta_{2}=0.5$ and a periodicity of $\tau\equiv\left(1+\sqrt{5}\right)/2$
. Bulk bands are clearly observed with their position remaining relatively
unchanged as $\delta$ is varied. Due to the presence of chiral symmetry
a topological insulator phase is verified, with persistent topological
zero-energy mode across all $\delta$ values, as well as edge-states
located in the gaps. (b) Spectrum of a physical realization of the
cAA model, corresponding to the phasonic parameter $\delta=0.2$.
The doubly degenerate zero-energy modes are highlighted in red, while
the additional visible edge states are marked in green. Notably, not
all finite energy edge states are visible for a particular $\delta$.
(c) Wave-functions for the modes highlighted in red. (d) Wave-functions
for the modes highlighted in green. \label{fig:Spectrum-and-Edge}}
\end{figure}

\end{widetext}

The spectral properties of the cAA model exhibit the remarkable complexity
expected from a quasicrystal: In the case where the hopping modulation
is commensurate with the lattice spacing, with $L=p/q\in\mathbb{Q}$,
the spectrum comprises $q$ well-defined bands, and the Brillouin
zone becomes, itself, $q-$periodic. However, in the quasicrystalline
limit where $L\notin\mathbb{Q}$, the spectrum transforms into a Cantor
set, lacking a well-defined band structure. Nevertheless, in this
latter case, the spectrum separates into regions characterized by
a dense arrangement of states in the thermodynamic limit, called quasibands.
Fig. \ref{fig:Spectrum-and-Edge} depicts a chain with $N=300$ lattice
sites, showcasing the effect incommensurate hopping modulations.
In particular, for this Figure, we take the periodicity to be equal
to the golden ratio $L=\tau$.

This Figure clearly illustrates the appearance of
quasibands, but also the existence of in-gap edge states in the cAA
model. The edge states connect between quasibands as the phasonic
degree of freedom is shifted, as demonstrated in panel (a)
of the Figure.

From the 1D point of view, this points to the ability
of sending zero-energy edge states into the bulk by manipulating the
phasonic degree of freedom $\delta$. Indeed, one important feature
to consider, is that a physical realization of the cAA model comprises
a single value of $\delta$ (see panel (b)). One should therefore
not interpret our figures such as panel (a) as physical ribbons of
material. Rather, only from the superspace interpretation of the phasonic
degree of freedom $\delta$ as a momentum along a synthetic direction
can one map the 1D chain into a ribbon like structure with periodic
boundary conditions along the synthetic direction. The physically
accessible spectrum is, in reality, only a slice of the ribbon spectrum
at a fixed value of $\delta$. This way, a realization of the 2D superspace
model actually corresponds to many possible physical realizations
of the cAA model, each related to each other by translations in the
phason $\delta$.

\section{Genesis of Dirac cones in the vanishing field limit}

To shed some light on the origin of the topological origin of the
edge state structure and further spectral properties of the cAA model,
as well as set up the necessary tools for the topological analysis
of the model. We now focus on what we call a vanishing field (small
$b)$ limit, or equivalently on a large periodicity (large $L$) limit.
To order $\mathcal{O}(b)$, the field dependence drops out from the
Hamiltonian, however, we can nevertheless consider its extension to
a superspace model. It is worth noting that this approximation must
be performed carefully, but it remains valid so long as the periodicity
is much larger than the total length of the system $N$, i.e. this
approximation can be preformed, for $b=p/q$, only in the limit $pN\ll q$.
The case of an incommensurate periodicity can be recovered trivially
by considering successive approximants $p/q$ to the desired value
of $b$. The advantage of this vanishing field treatment is that many
of the model's properties can be analytically computed in this limit,
and intuition can be gained and later utilized for the generic case
of finite and possibly large fields. For vanishing fields, the Hamiltonian
reads, in a real space description

\begin{align}
H_{\text{2D}}= & -\sum_{i,j}t_{1}a_{i,j}^{\dagger}b_{i,j}+\frac{\Delta_{1}}{2}a_{i,j}^{\dagger}b_{i,j+1}+\text{h.c.}\nonumber \\
 & -\sum_{i,j}t_{1}b_{i,j}^{\dagger}a_{i+1,j}+\frac{\Delta_{2}}{2}b_{i,j}^{\dagger}a_{i+1,j+1}+\text{h.c.},\label{eq:2Dreal_space}
\end{align}
where we have switch from the variable $n$ indexing physical state
sites to the pair $\left(i,j\right)$, where $j$ represents the lattice
index along the synthetic dimension. The change from $n$ to $i$
is merely a notational convention to place both directions in the
same footing. One of the advantages of this limit is that we can analyze
the case where we have an infinitely long chain with periodic boundary
conditions, such that it becomes possible to construct Brillouin zone
parameterized by momenta along the synthetic direction $\delta\in\left[0,2\pi\right[$
and the physical direction $k\in\left[0,2\pi\right[$. Denoting $\boldsymbol{k}=\left(k,\delta\right)$
also allows for a representation of the superspace Hamiltonian in
the form of a matrix $H_{2D}(\boldsymbol{k})$ as $H_{\text{2D}}=\sum_{\boldsymbol{k}}\Psi_{\boldsymbol{k}}^{\dagger}H_{2D}(\boldsymbol{k})\Psi_{\boldsymbol{k}}$,
where $\Psi_{\boldsymbol{k}}^{\dagger}=\left(a_{\boldsymbol{k}}^{\dagger},b_{\boldsymbol{k}}^{\dagger}\right)$,
and with $H_{2D}(\boldsymbol{k})$ explicitly given by (see Appendix
A for details)

\begin{align}
H_{2D}(\boldsymbol{k})= & -\left[t_{1}+\Delta_{1}\cos\delta+\left(t_{2}+\Delta_{2}\cos\delta\right)\cos k\right]\sigma_{x}\nonumber \\
 & -\left[\left(t_{2}+\Delta_{2}\cos\delta\right)\sin k\right]\sigma_{y}.\label{eq:HamiltDirac}
\end{align}
This shows that the vanishing field Hamiltonian can be realized in
the form of a Dirac model, at the cost of the $\sigma_{x}$ and $\sigma_{y}$
Pauli matrices, but does not include a $\sigma_{z}$. This is to be
expected, once again, due to the presence of chiral symmetry. Indeed,
we see that coupling between replicas of the model along $j$ preserves
the fact that hopping can occur only between $A$ and $B$ sublattices
or vice-versa. Again, if one were to consider an
AA type on-site modulation, this would induce couplings among different
sites of the $A$ sublattice and of the $B$ sublattice, thus breaking
the chiral symmetry of the 1D chain. Now, for all quasicrystalline
systems, a treatment using synthetic dimensions is possible, however,
chiral symmetry makes this model extremely interesting, since the
superspace model becomes a semimetal with topological properties.

We say that it is a semimetal since the cAA model can host at most
$4$ distinct massless Dirac cones in this vanishing field limit,
the positions of which are not restricted to any high-symmetry points
of the Brillouin zone, but rather can wander around certain lines
within it, depending on the parameter space spanned by $t_{1},t_{2},\Delta_{1}$
and $\Delta_{2}$. To show that this is the case, given Equation (\ref{eq:HamiltDirac}),
it is simple to diagonalize the Hamiltonian and
find the energy spectrum $\varepsilon(k,\delta)$. Finding the points
at which massless Dirac cones occur, then becomes a matter of analytically
solving $\varepsilon(k,\delta)=0$, which yields the solutions

\begin{align}
\delta_{D}(k) & =\pm\arccos\left(\frac{e^{ik}t_{1}+t_{2}}{e^{ik}\Delta_{1}+\Delta_{2}}\right)\cap\mathbb{R}\nonumber \\
 & \lor\pm\arccos\left(\frac{t_{1}+e^{ik}t_{2}}{\Delta_{1}+e^{ik}\Delta_{2}}\right)\cap\mathbb{R}.\label{eq:Dirac_cones}
\end{align}
We have explicitly included an intersection with the real numbers
to make explicit the fact that the $\arccos(z)$ function may have
solutions corresponding to complex numbers, however we retain only
real solutions, corresponding to real momenta $\delta$. Additionally,
near the location of these Dirac cones, the energy dispersion can
be extracted analytically. If we expand the momentum radially near
$\boldsymbol{K}_{D}=\left(k,\delta_{D}(k)\right)$ as $\boldsymbol{q}=\boldsymbol{K}_{D}+q\left(\cos\theta,\sin\theta\right)$
with $q=\sqrt{k^{2}+\delta^{2}}$, we can write down a linearized
form of the dispersion. In particular, if we consider the terms
\begin{align}
\eta_{1} & =\left(t_{2}\Delta_{1}-t_{1}\Delta_{2}\right)^{2}/\left(\Delta_{1}-\Delta_{2}\right)^{2},\\
\eta_{2} & =\left(t_{2}\Delta_{1}-t_{1}\Delta_{2}\right)^{2}/\left(\Delta_{1}+\Delta_{2}\right)^{2},\\
\xi_{1} & =\left(t_{1}-t_{2}+\Delta_{1}-\Delta_{2}\right)\left(t_{2}-t_{1}+\Delta_{1}-\Delta_{2}\right),\\
\xi_{2} & =\left(\Delta_{1}+\Delta_{2}-t_{1}-t_{2}\right)\left(t_{1}+t_{2}+\Delta_{1}+\Delta_{2}\right),
\end{align}
then the linearized dispersion around the Dirac cones reads

\begin{align}
\varepsilon_{1} & =\pm q\sqrt{\eta_{1}\cos^{2}\theta+\xi_{1}\sin^{2}\theta},\label{eq:Dirac_cone_Aniso1}\\
\varepsilon_{2} & =\pm q\sqrt{\eta_{2}\cos^{2}\theta+\xi_{2}\sin^{2}\theta}.\label{eq:Dirac_cone_Aniso2}
\end{align}
Evidently, these expressions hold only when the massless Dirac cones
exist, i.e. when there exists a solution to Eq. (\ref{eq:Dirac_cones}).
As can be seen from (\ref{eq:Dirac_cone_Aniso1}) and (\ref{eq:Dirac_cone_Aniso2}),
these Dirac cones are generically anisotropic since $\eta_{i}\neq\xi_{i}$.
By probing the parameter space of the model, we identify five possible
phases for the superspace model: Phase (i) corresponds to a semimetal
phase with two massless Dirac cones at $k=0$; Phase (ii) corresponds
to a similar semimetal phase with two massless Dirac cones at $k=\pi$;
Phase (iii) is an insulating phase, with no massless Dirac cones;
Phase (iv) is a semimetallic phase, with four massless Dirac cones,
two of which are located at $k=0$ and another two at $k=\pi$. For
all phases where Dirac cones are visible, they are arranged symmetrically
around $\delta=0$. Phase transitions between these four phases occur
due to merging and subsequent lifting of the Dirac cones as the parameters
of the model are tuned. These mergings can occur either at the edge
($\delta=\pm\pi)$ or the center $(\delta=0)$ of the Brillouin zone
along $\delta$. The final phase (v) occurs only for very specific
choices of the parameters, where at the critical point of a simultaneous
fusion of Dirac cones at $k=0$ and $k=\pi$ a nodal line semimetal
phase is observed, with the two bands touching along all values of
$k$ and $\delta=0$. 

\begin{figure}
\begin{centering}
\includegraphics[scale=0.23]{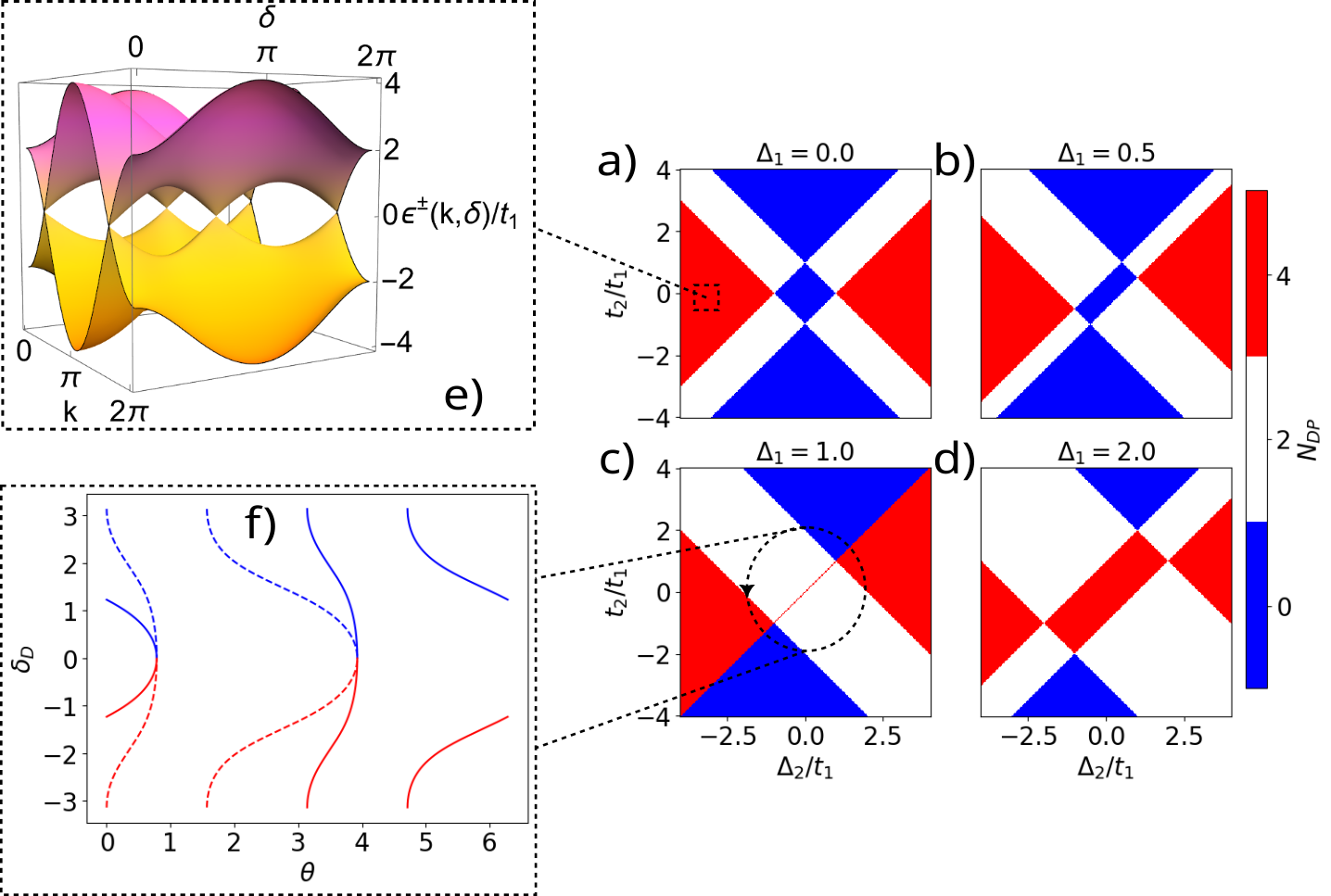}
\par\end{centering}
\caption{(a), (b), (c) and (d): Phase diagrams for the number of Dirac points
$N_{DP}$ as a function of the models parameters measured relative
to $t_{1}$. In (c), the line $\Delta_{2}=t_{2}$ corresponds to a
situation where the bands touch not only at specific points, but rather
along a full line, at $\delta=0$ and for any value of $k$, such
that the model actually is in a topological nodal line semimetal phase.
(e) Band structure of the cAA model showcasing the existence of 4
massless Dirac cones occurring for the parameters $t_{1}=1,t_{2}=0,\Delta_{1}=0,\Delta_{2}=-3$.
(f) Trajectories of the Dirac cones for $k=0$ (continuous lines)
and $k=\pi$ (dashed lines), as a circle is traversed in the parameter
space as indicated in (c). Note that Dirac cones always come in pairs
located one above (blue) and one below (red) the line $\delta=0$.
The hopping amplitudes are parameterized with $\Delta_{2}=2\cos\theta$
and $t_{2}=2\sin\theta$, and $\theta$ is varied from 0 to $2\pi$.
Several Dirac cone mergers are observed as $\theta$ is varied both
at $k=0$ and $k=\pi$.\label{fig:Dirac-Locations-1}}
\end{figure}

We illustrate the wandering, merging and subsequent appearance or
and disappearance of massless Dirac cones in Fig. \ref{fig:Dirac-Locations-1}(f)
for a specific choice of parameters. Different phase diagrams can
also be drawn by picking three parameters with varying strength relative
to a fourth one. For instance, we can vary $\Delta_{2}/t_{1}$ and
$t_{2}/t_{1}$ while choosing a few different values of $\Delta_{1}/t_{1}$.
For values of $\Delta_{1}/t_{1}=0,0.5,1,2$, we plot some phase diagrams
in Fig. \ref{fig:Dirac-Locations-1}.

Remarkably, the number of pairs of Dirac cones in this model can be
shown to be related to the topological properties of the 1D model.
To see why this is the case, we can think of the momentum $\delta$
as a tunable parameter, which is likely indeed be the case for any
realistic implementation of the cAA model. If this is so, a massless
Dirac cone occurring in the model corresponds to the band gap closing
when $\boldsymbol{K}_{D}=\left(k,\delta_{D}(k)\right)$ is approached.
For this reason, it becomes very useful to consider a winding number
as a function of $\delta$, such that each particular physical implementation
of the cAA model has its own $W(\delta)$. Due to the presence of
chiral symmetry, the Hamiltonian matrix can be brought into the form

\begin{equation}
H(\boldsymbol{k})=\begin{bmatrix}0 & q(\boldsymbol{k})\\
q^{*}(\boldsymbol{k}) & 0
\end{bmatrix},
\end{equation}
and the Winding number can be defined as

\begin{equation}
W(\delta)=\int_{0}^{2\pi}dk\frac{\partial}{\partial k}\log q(k,\delta).\label{eq:WInt}
\end{equation}
The reason we may be interested in such a quantity, besides the fact
that it signals the presence or absence of a zero-energy edge mode
at each $\delta$ is the fact that points at which a transition between
phases with $W(\delta)=1$ and $W(\delta)=0$. These are precisely
the set of points at which massless Dirac cones occur. An analytical
computation of the winding number from Equation (\ref{eq:WInt}),
results in Equation (\ref{eq:Winding_result}) 

\begin{equation}
W(\delta)=\Theta\left(\left|t_{1}+\Delta_{1}\cos\delta\right|-\left|t_{2}+\Delta_{2}\cos\delta\right|\right),\label{eq:Winding_result}
\end{equation}
as could be anticipated from the topological properties of a standard
SSH model. Here, $\Theta(x)$ represents the Heaviside step function
of $x$. 

Now, the parity number of pairs of Dirac points, $N_{2DP}$mod 2,
can be counted by looking at whether the Winding number is the same
when $\delta=0$ and when $\delta=\pi$. One considers, in particular,
the Winding number at $\delta=0,$ since the Dirac cones always come
in pairs, due to the Fermion doubling theorem \citep{Nielsen1981,Nielsen1981_2},
and furthermore are always located symmetrically around this point.
Hence, we know that if the Winding number is the same at $\delta=0$
and $\delta=\pi$ then it must have changed an even number of times
between these two values of $\delta$, leading to an even number of
pairs of Dirac points. Otherwise, if it is not the same, it must have
changed an odd amount of times, leading to an odd number of pairs
of Dirac points. Hence, we have the simple formula of Equation (\ref{eq:N2PDP}).

\begin{align}
N_{2DP}\text{mod }2 & =|W(\pi)-W(0)|.\label{eq:N2PDP}
\end{align}
This quantity already provides some information regarding the presence
or absence of masless Dirac cones, but the total number of of Dirac
cones $N_{DP}$ in the Brillouin zone cannot be counted in this manner.
One simple way to augment this result is to sum over changes in $W(\delta)$
along the entire range of $\delta$, instead of considering only $\delta=0$
and $\delta=\pi$. In particular, since the winding number varies
as a series of step-functions, taking the derivative of $W(\delta)$
along $\delta$ will yield a series of Dirac-delta functions. If these
delta functions are integrated over, each one contributes with unity,
and thus we can count the number of times such changes occur, and
thus we have the equation (\ref{eq:NDP}).

\begin{equation}
N_{DP}=\int_{0}^{2\pi}d\delta\left|\frac{\partial}{\partial\delta}W\left(\delta\right)\right|.\label{eq:NDP}
\end{equation}
These considerations can also be applied to ribbons of familiar 2D
materials such as graphene, and we illustrate some analogies between
the cAA model and zig-zag ribbons of graphene in Appendix B, providing
a topological description of the emergence of zero energy edge states
in these configurations. 

For now, in order to demonstrate the practical application of the
aforementioned formula (\ref{eq:NDP}), we have employed it to compute
phase diagrams for the model, as depicted in Figure \ref{fig:Dirac-Locations-1}
(a)-(d). The diagrams clearly indicate that the count of Dirac cones
consistently falls into one of three possibilities: 0, 2, or 4, regardless
of the model's parameters. Indeed, these diagrams, along with Fig.
\ref{fig:Dirac-Locations-1} (f) make it clear that it becomes possible
to induce a transition from a state characterized by the presence
of an energy band gap to one featuring a gapless state with Dirac
cones. The genesis of Dirac cones will have observable effects, even
when the system is projected down to physical space, particularly
concerning edge states in the 1D physical model when the AA hopping
modulation is increased. A detailed exploration of this phenomena
is presented in the following section. 

\section{Coexistence of 1D and 2D topology at finite fields}

To study the effects of a magnetic field on the superspace 2D model
in more detail, i.e. if one moves away from the limit of slowly varying
hopping modulations for the 1D physical system, a number of possible
approaches present themselves, but none is as clean as the analytic
treatment provided in the previous section. The added difficulty stems
from the explicit breaking of translational symmetry of the 1D model.
The standard toolkit for the IQHE could be applied by analyzing the
$L\in\mathbb{Q}$ case and the notion of the magnetic translation
group, however, a simple analysis of an incommensurate modulation
remains, nonetheless analytically out of reach. For this reason, we
implement a numerical method based on recently proposed topological
markers for Dirac models and quasicrystals which is capable of handling
the problem in full. Two topological markers become relevant for our
discussion: The first is a winding number marker and the second is
a quasicrystal Chern marker. Since the cAA is a Dirac model, both
are ``different dimensional shadows'' of the same invariant known
as the wrapping number as pointed out by von Gersdorff and Chen \citep{Gersdorff_2021},
however, we take different approaches to their calculation, and use
the method of Chen \citep{Chen_2022} only for the winding number
marker. This approach is based on the construction of a 1D universal
topological operator of the form

\begin{equation}
\hat{C}_{1D}(\delta)=N_{D}\mathcal{W}\left[Q(\delta)\hat{x}P(\delta)+P(\delta)\hat{x}Q(\delta)\right],\label{eq:TopMarker}
\end{equation}
where generically $N_{D}$ is a normalization factor, here simply
unity, and $\mathcal{W}$ is a Pauli matrix which remains unused in
the Dirac model Hamiltonian: the $\sigma_{z}$ Pauli matrix. $P$
and $Q$ are projectors into occupied and unoccupied states.

\begin{align}
P & =\sum_{E<E_{F}}\left|n_{E}\right\rangle \left\langle n_{E}\right|,\label{eq:Proj_Real_space}\\
Q & =\mathds{1}-P.
\end{align}
For clarity, we make here the assumption that the cAA model is fermionic,
which on the one hand makes the concept of occupation of bands well
defined, and on the other hand, sets the IQHE interpretation of the
superspace description on firmer ground, since this effect is often
considered to be reserved for fermionic models. Although this is not
really the case, with Hofstadter models having been obtained for bosonic
systems as well \citep{Krivosenko2021}, we nevertheless stick to
the analysis of fermionic systems for clarity. Furthermore, expected
implementations in cold-atom systems or spin-systems rely on fermions
(electrons and Jordan-Wigner Fermions respectively). The use of fermions
allows for the consideration of a Fermi energy $E_{F}$, and for all
future discussion and computations presented in this paper, we will
focus on neutral systems, with $E_{F}=0$. Finally, in Equation (\ref{eq:TopMarker}),
$\hat{x}$ is the position operator, which, as stated in Section \ref{sec:The-Chiral-Aubry-Andr=0000E9},
is assumed to have the same eigenvalue for states localized within
each of the two sites within the unit cell of the cAA model. The local
winding number marker at a given position $x$ of the model is then
generically computed as the diagonal matrix element of the topological
operator

\begin{equation}
W(x,\delta)=\left\langle x\right|\hat{C}_{1D}(\delta)\left|x\right\rangle =\sum_{\sigma}\left\langle x,\sigma\right|\hat{C}_{1D}(\delta)\left|x,\sigma\right\rangle ,
\end{equation}
where $\sigma$ indicates, generically, all internal degrees of freedom
of the model. In the BDI symmetry class, for the SSH model, $\sigma=A,B$
are simply the two pseudo-spin projections (sublattice types). Indeed,
the Hamiltonian $H_{\text{odAA}}(\delta)$ (Eq. \ref{eq:odAA}) is
here interpreted simply as a disorder term, and hence, even in the
case where the unit cell becomes effectively larger due to the periodicity
of modulation, we always retain the summation to be over the two sites
corresponding to the unit cells of the model described by $H_{\text{SHH}}$
(Eq. \ref{eq:SSH}). When we consider a finite cAA chain, this winding
number marker is expected to coincide with the winding number topological
invariant deep in the bulk, and decay near its boundaries. For this
reason, due to the presence of spatial disorder, and for numerical
stability, an averaging over a finite region $\mathcal{R}$ of size
$N_{\mathcal{R}}$ deep in the bulk is performed, so that the behavior
at the edges is avoided. In the end, the winding number invariant
can be computed as

\begin{equation}
W(\delta)=\frac{1}{N_{\mathcal{R}}}\text{tr}_{\mathcal{R}}\left(\hat{C}_{1D}(\delta)\right)=\frac{1}{N_{\mathcal{R}}}\sum_{x\in\mathcal{R}}W(x,\delta).
\end{equation}
Note that, in the cAA model, this topological invariant is computed
purely from a single slice of the superspace model at a fixed $\delta$,
and, once again, the role played by the odAA term is merely one of
introducing spatial disorder. On the other hand, to access the Chern
number, the phasonic degree of freedom must come into play, and all
slices of fixed $\delta$ become relevant. The use of a 2D universal
topological marker as proposed by Chen, however, would be unwise,
because the cAA model is actually only disordered in the physical
1D space. A 2D universal topological marker is much better suited
to the analysis of disordered 2D lattices, and thus, as an alternative,
a recently proposed method originally due to Sykes and Barnett \citep{Sykes_2022}
is therefore employed. This method is particularly suited for handling
models like the AA models or other 1D quasicrystals, and its underlying
insight is to interpret the phasonic degree of freedom $\delta$ as
a function of time, $\delta(t)=2\pi t/T$, which enables us to think
of the 1D chain as undergoing an adiabatic and $T-$periodic evolution.
This interpretation is made physical in systems exhibiting the so-called
Thouless or topological pump, which is a quantized pump of charge
from one edge of a 1D chain to the other over the course of a period
$T$ \citep{Thouless_1983}. The topological pump
can be directly observed when a 1D crystal is periodically driven
by an external field. The charge pump is topologically protected by
what is effectively the 2D Chern number for the superspace model.

By interpreting the phasonic degree of freedom as
proportional to time, regardless of its actual nature (in previous
discussions we have assumed that $\delta$ is a fixed parameter),
we can compute the topological pump due to an effective adiabatic
Hamiltonian \citep{Sykes_2022}, which can be shown to be

\begin{equation}
h(t)=i\left[\frac{\partial P(t)}{\partial t},P(t)\right],\label{eq:heff}
\end{equation}
This effective Hamiltonian induces an effective adiabatic evolution
operator $U(t)=\exp\left(iht\right)$, and $P(t)=U^{\dagger}(t)PU(t)$
is the projector onto occupied states at time $t$. The local topological
quasicrystal marker \citep{Sykes_2022} can, in turn, be defined as
the expectation value of a topological operator $\hat{M}_{1Q}(\delta)$,
as in

\begin{align}
M_{1Q}(x,\delta) & =\left\langle x\right|\hat{M}_{1Q}(\delta)\left|x\right\rangle \nonumber \\
 & \equiv\left\langle x\right|P(t)U^{\dagger}(t)\hat{x}U(t)P(t)\left|x\right\rangle .
\end{align}
Much like in the case of the winding number marker, averaging over
a region $\mathcal{R}$ deep in the bulk of the quasicrystal is necessary,
and we compute $M_{1Q}(t)=\text{tr}_{\mathcal{R}}\left(\hat{M}_{1Q}(t)\right)/L_{\mathcal{R}}$.
This quantity measures the location of the so-called Wannier center
at time $t$, and, as described by the modern theory of Polarization
\citep{Vanderbilt_2018,Spaldin_2012} (or the Thouless pump). Then,
the difference in position of the Wannier center at time $t_{0}$
and $t_{0}+T$ gives the polarization change or charge transfer over
the course of a period of the driving. This charge transfer is nothing
but the 2D Chern number characterizing the superspace model, and hence,
setting $t_{0}=0$ for simplicity, at which point $U(0)=\mathds{1}$,
the invariant is \citep{Sykes_2022}

\begin{equation}
C=M_{1Q}(T)-M_{1Q}(0).\label{eq:ChernMarker}
\end{equation}
Armed with these two topological invariants, $C$ and $W(\delta)$,
and the respective methods of computation in real space, we are primed
to characterized the topology of the cAA model. Note, that the winding
number characterizes the topology of the model for each value of $\delta$,
and hence, for a computation of $C$ one may actually be integrating
over topological phases with $W(\delta)=0$ and $W(\delta)=1$, which
is another pointer to the subtle nature of how $C$ characterizes
the 1D phase, despite being an invariant which finds a natural setting
in 2D. 

A simple way to illustrate the power of the quasicrystal topological
marker is to compute the Chern number for the cAA model as a function
of the modulation periodicity and Fermi energy of the model. For each
value of $b$ and $E_{F}$, we compute $M_{1Q}(0)=\text{tr}_{\mathcal{R}}\left(P(0)\hat{x}U(t)P(0)\right)/N_{\mathcal{R}}$,
and afterwards compute the time evolved projection operator $P(t)$
by slicing the period $T$ into slices of size $\Delta t$. 

\begin{figure}
\begin{centering}
\includegraphics[scale=0.15]{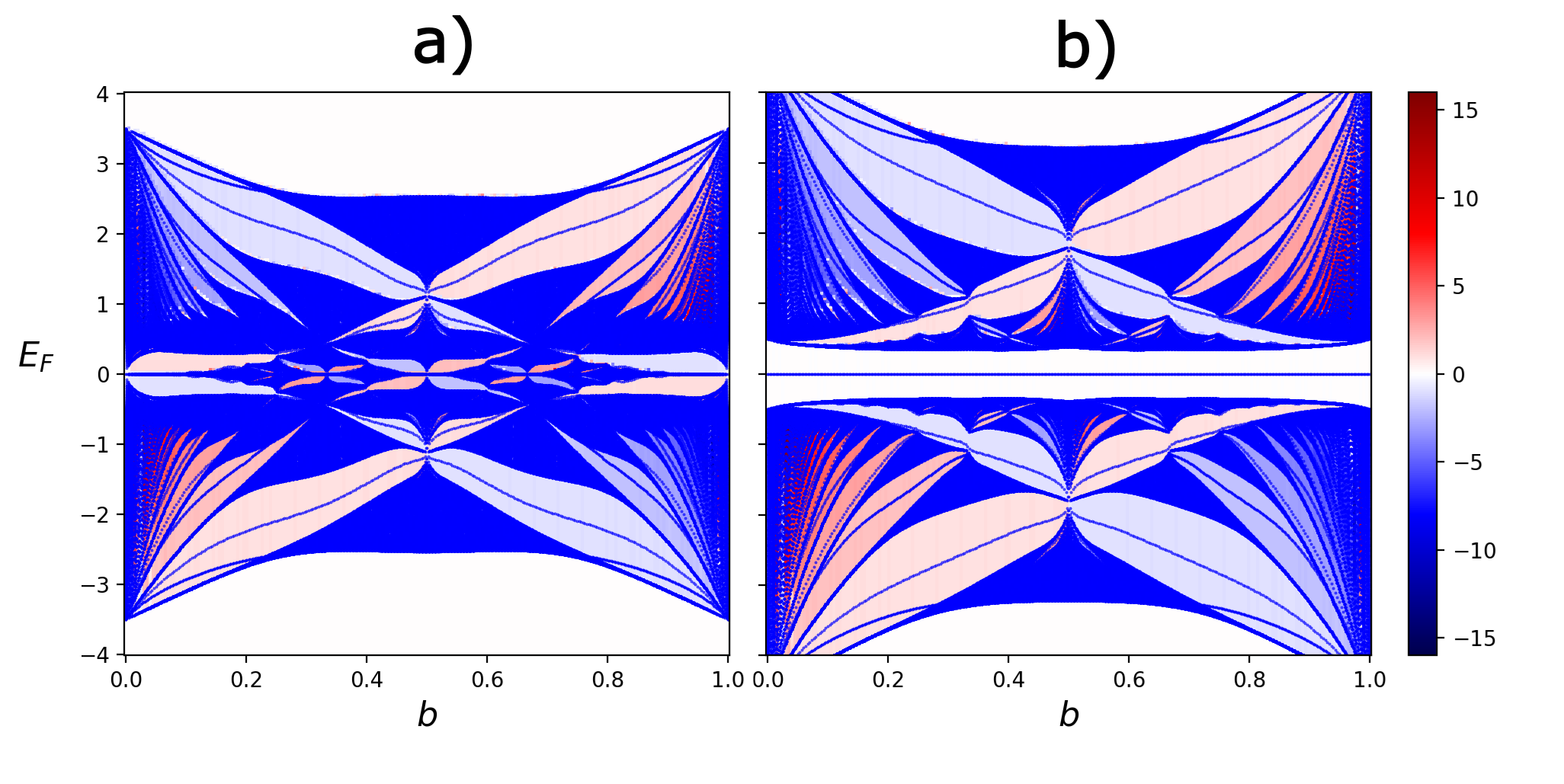}
\par\end{centering}
\caption{Chern number calculated from the topological marker as a function
of the periodicity $b$ and the Fermi Energy $E_{F}$. The resulting
plots produce a fractal image corresponding to a colored Hofstadter
butterfly, where we the coloring indicates the value of the Chern
number, and the thick blue lines indicate the energy levels. (a) Semimetal
phase, with parameters $t_{1}=t_{2}=\Delta_{1}=1$ and $\Delta_{2}=1.5$.
At charge neutrality, there is no trivial Chern phase, and the Chern
number jumps from $C=-1$ to $C=1$ or vice-versa. This is analogous
to the case of the half-integer quantum hall effect in graphene. (b)
Insulating phase with $t_{1}=t_{2}=\Delta_{1}=1$ and $\Delta_{2}=0.5$.
At neutrality, there exists a trivial Chern phase. This is analogous
to the integer quantum hall effect observed, for instance a square
lattice material. The presence of the zero-energy modes in panel (b)
is indicative of the presence of non-trivial 1D topology, accounted
for by a winding number $W(\delta)=1$. Although this is not so clearly
visible in panel (a), these states can also exist at zero energy for
semimetal phases at certain values of $\delta$, but are generically
not present for all values of modulation shift. \label{fig:Chern-number-calculated}}
\end{figure}

A $U(t+\Delta t)\approx\exp\left(-ih(t)\Delta t\right)U(t)$ at each
successive interval, one can calculate the Wannier center $M_{1Q}(t)$,
and iterating this procedure over one period, we can compute $C$
using (\ref{eq:ChernMarker}). Computing $C$ for each value of $b$
and $E_{F}$ reveals a striking illustration of a colored Hofstadter
butterfly \citep{Hofstadter_1976}. Two examples of these colored
Hofstadter butterflies, for topological insulating and topological
semimetal phases are provided in Fig. \ref{fig:Chern-number-calculated}.

Furthermore, the winding number topological marker clearly counts,
for each value $\delta$, the presence or absence of the zero-energy
edge mode even when $b$ is a finite quantity. In particular, this
enables us to count the number of Dirac cones using Eq. \ref{eq:NDP},
in exact analogy to the approach we took for vanishing fields. A band
touching point induces an abrupt change in $W(\delta)$, even when
the modulation periodicity is irrational. 

\begin{figure}
\begin{centering}
\includegraphics[width=0.45\textwidth]{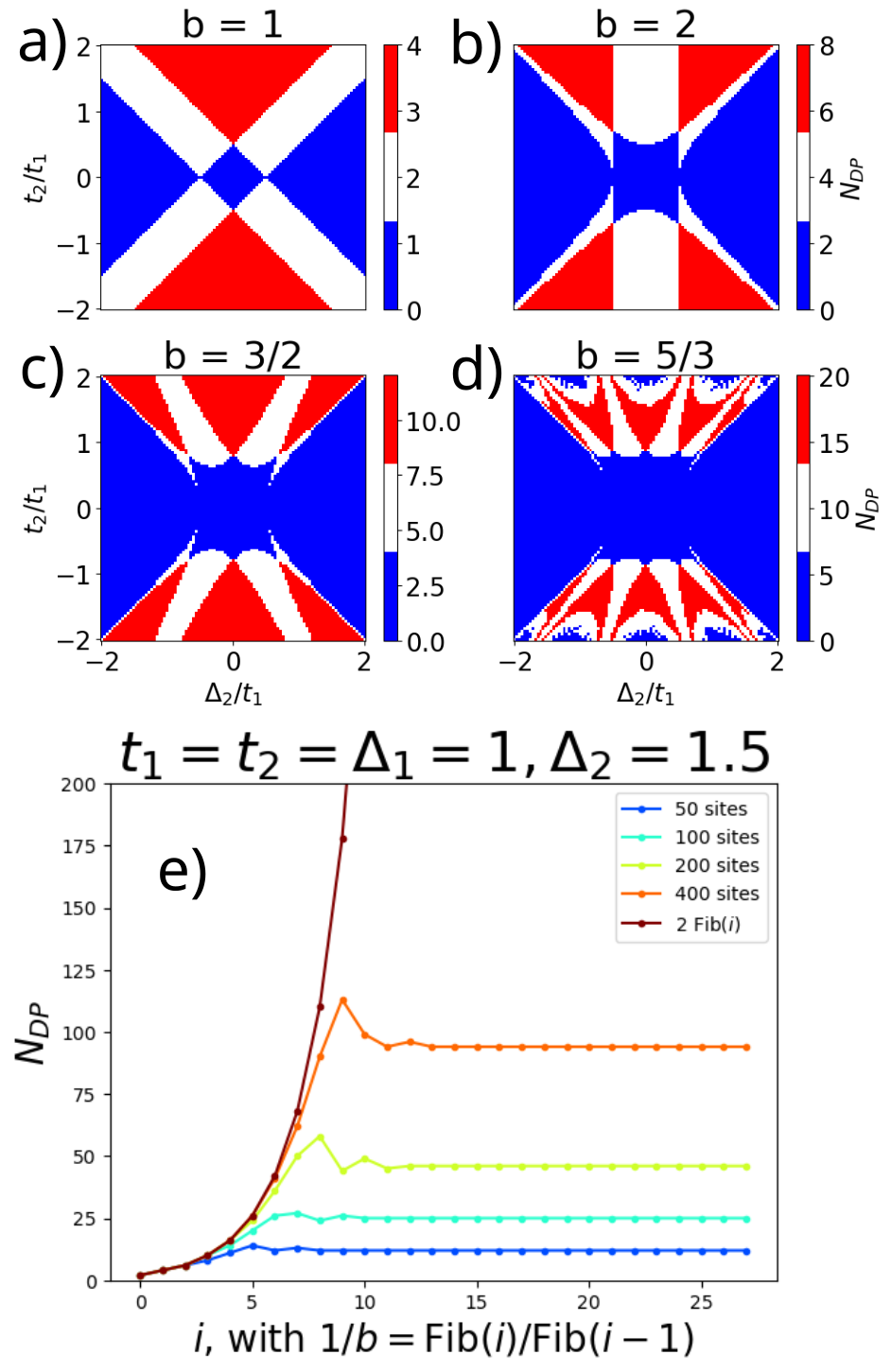}
\par\end{centering}
\caption{(a)-(d). Number of Dirac nodes calculated from our method relying
on the 1D winding number for finite $b$ and fixed $\Delta_{1}=0$.
For each of the phase diagrams, successive approximants to the golden
ratio $\tau=\left(1+\sqrt{5}\right)/2$ are utilized as the inverse
field. Explicitly, we use $b=1,2,3/2$ and $5/3$. (e) Growth of the
number of nodes as better approximants are utilized. The colored curves
correspond to different system sizes. Increasing this size shifts
the point at which the number of Dirac nodes stabilizes. We compare
the curves which give the number of Dirac cones with the four times
the $i$th Fibonacci number, i.e. with $2\text{Fib}(i)=2q$. We see
that the behavior of the curves saturates near the size of the system,
i.e. for $2q\sim N$, and begins deviating from the expected behavior
slightly earlier, when finite size effects come into play and already
a few unit cells become comparable with the system size. \label{fig:Approaching_Irrational_Period}}
\end{figure}

Let us consider the example case where $b$ is given
by successive rational approximations to the inverse golden ratio
$1/\tau$. Numerically, as $1/\tau$ is approached, the method becomes
more costly due to the necessity of a large sampling of points along
$\delta$, but nonetheless we can illustrate the behavior of convergence
to an irrational periodicity by producing several plots of phase diagrams
depending on the model's parameters. These are presented in Fig. \ref{fig:Approaching_Irrational_Period}
(a)-(d). We observe that, as the irrational limit is approached, the
number of massless Dirac cones increases. In particular for finite
and rational values $1/b=p/q$ of the field, the number of massless
Dirac cones appears to always be given by $0,2q$ or $4q$. 

For infinite chains, we would expect this behavior
to continue ad infininitum, with the bands becoming perfectly flat
and gapless in the limit of infinite unit cell and system size, as
the Dirac cones become denser in the Brillouin zone. Furthermore,
as the periodicity is brought closer and closer to an irrational value,
or in other words, if we focus on the large $q$ limit and keep $p/q$
finite, the system's physical size can also become relevant, leading
to a departure from the seemingly observed trend of the number of
$0,2q$ or $4q$ Dirac cones, instead stabilizing at some finite value.
This is observed in Fig. \ref{fig:Approaching_Irrational_Period}
(e).

The crux of this discussion is that the genesis of
Dirac cones in the superspace model of the cAA is found to be mappeable
to changes in a topological invariant as the the phasonic degree of
freedom is spanned over. We find this to be an interesting result
in and of itself, as tuning $\delta$ could be a useful knob for tuning
physical properties such as the conductivity at half-filling.

Besides this, we can now use these topological markers to explicitly
showcase the coexistence of 1D and 2D topology in the cAA model, by
computing both invariants for some parameter values. This calculation
is performed in Fig. \ref{fig:Gap-Chern-numbers,}, where we find
phases of type $\left(C,W\right)=\left(0,1\right),\left(\pm1,1\right),\left(\pm2,1\right),\left(\pm3,1\right)$
and $\left(\pm1,0\right)$. As can be clearly seen in this Figure,
the Chern number counts the number of in-gap edge states for all possible
values of $\delta$. A point which cannot go without mentioning is
that even though a Chern number exists and this bulk-boundary correspondence
is verified in the superspace model, the number of 1D edge modes which
occur in a given energy gap of any particular physical realization
of the cAA model is not in direct correspondence with this Chern number.
The 2D topological invariant merely points to how many edge states
\textit{can} occur at the edge of the 1D model if the odAA modulation
is shifted relative to the lattice. 

Nevertheless, for a physical realization of the cAA model, and for
a Fermi energy $E_{F}$ located within a gap above the zero energy
$\varepsilon=0$, it is clear that both $W$ and $C$ are necessary
quantities to characterize the topology of the model, as zero energy
edge states will be measured depending on $W(\delta)$,
and finite energy edge states will be visible depending on both $C$
and $\delta$.

This coexistence can then be manifest in many different
ways, depending on whether the modulation is commensurate with the
lattice, incommensurate with the lattice, as well as depending on
the periodicity itself. Fig. \ref{fig:Gap-Chern-numbers,} showcases
both commensurate and incommensurate modulations, yielding in the
latter case a zero-energy edge mode visible for all $\delta$, and
in the former a zero-energy edge mode which wanders into the bulk
for some values of $\delta$. As can be seen from this Figure, the
position, number and distribution of gaps, as well as edge-states
within each gap, depends on the model's parameters, such as the relative
magnitudes of the $J_{i}$s and $\Delta_{i}$s. However, the evaluation
of $W(\delta)$ and $C$ is always possible, and provides a signature
or probe to the appearance of edge modes.

\begin{widetext} 

\begin{figure}[H]
\begin{centering}
\includegraphics[width=0.9\textwidth]{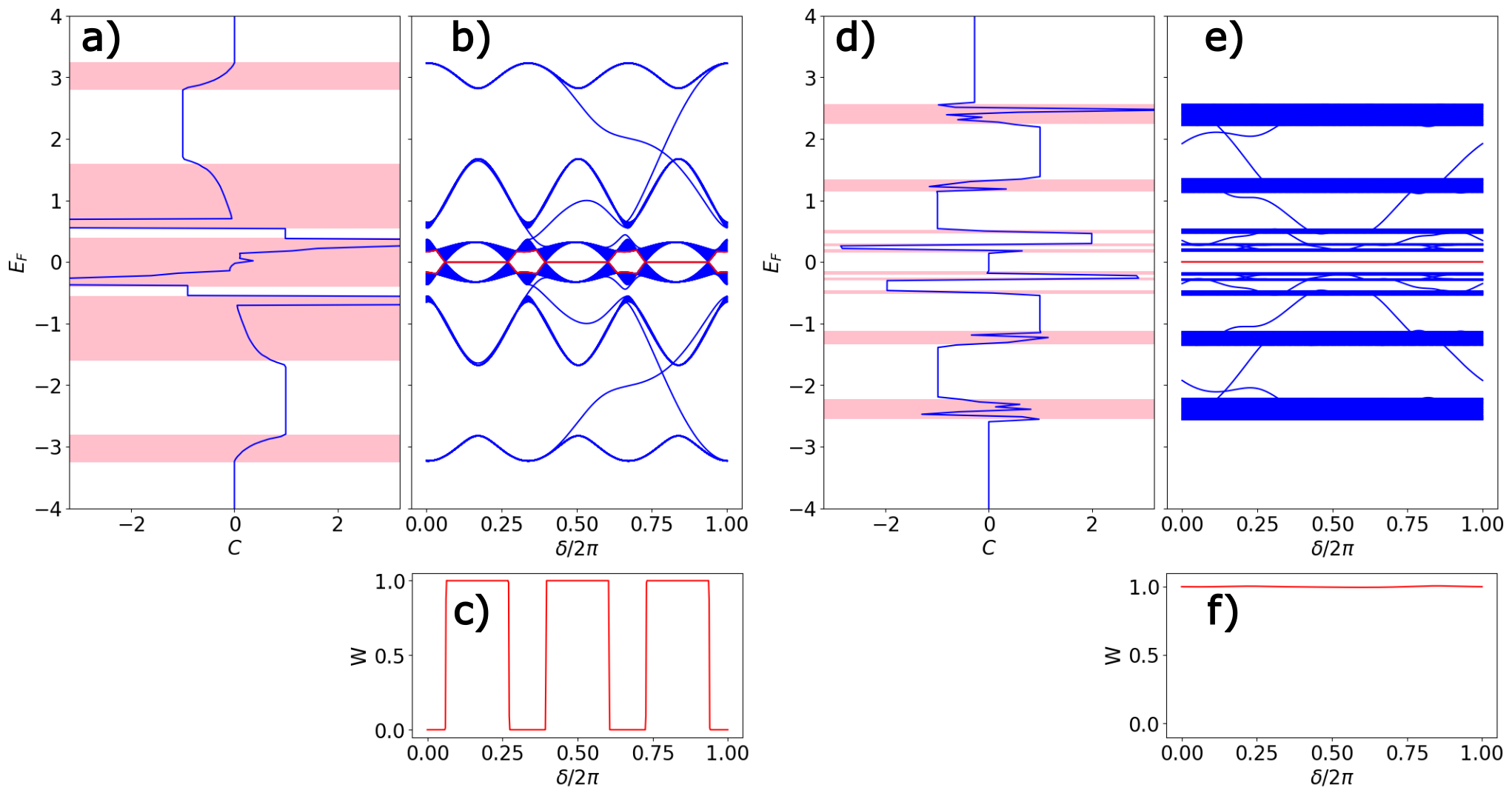}
\par\end{centering}
\caption{Gap Chern numbers, energy spectrum, and winding number for a cAA model
with a commensurate modulation ($b=1/3$ (a), (b) and (c) respectively),
and with an incommensurate modulation ($b=1/\tau)$ (d), (e) and (f)
respectively). Distinct topological phases are visible,
with invariants $\left(C,W\right)=\left(0,\pm1\right),\left(1,\pm1\right)$
in panels (a)-(c) as well as $\left(C,W\right)=\left(1,\pm1\right),\left(1,\pm2\right),\left(1,\pm3\right)$
in panels (d)-(f). The gap Chern numbers fluctuate within the regions
marked in pink, where bulk bands are present, and converge to finite
values within the bulk energy gaps.} \label{fig:Gap-Chern-numbers,}
\end{figure}

\end{widetext}

\begin{figure}
\begin{centering}
\includegraphics[width=0.45\textwidth]{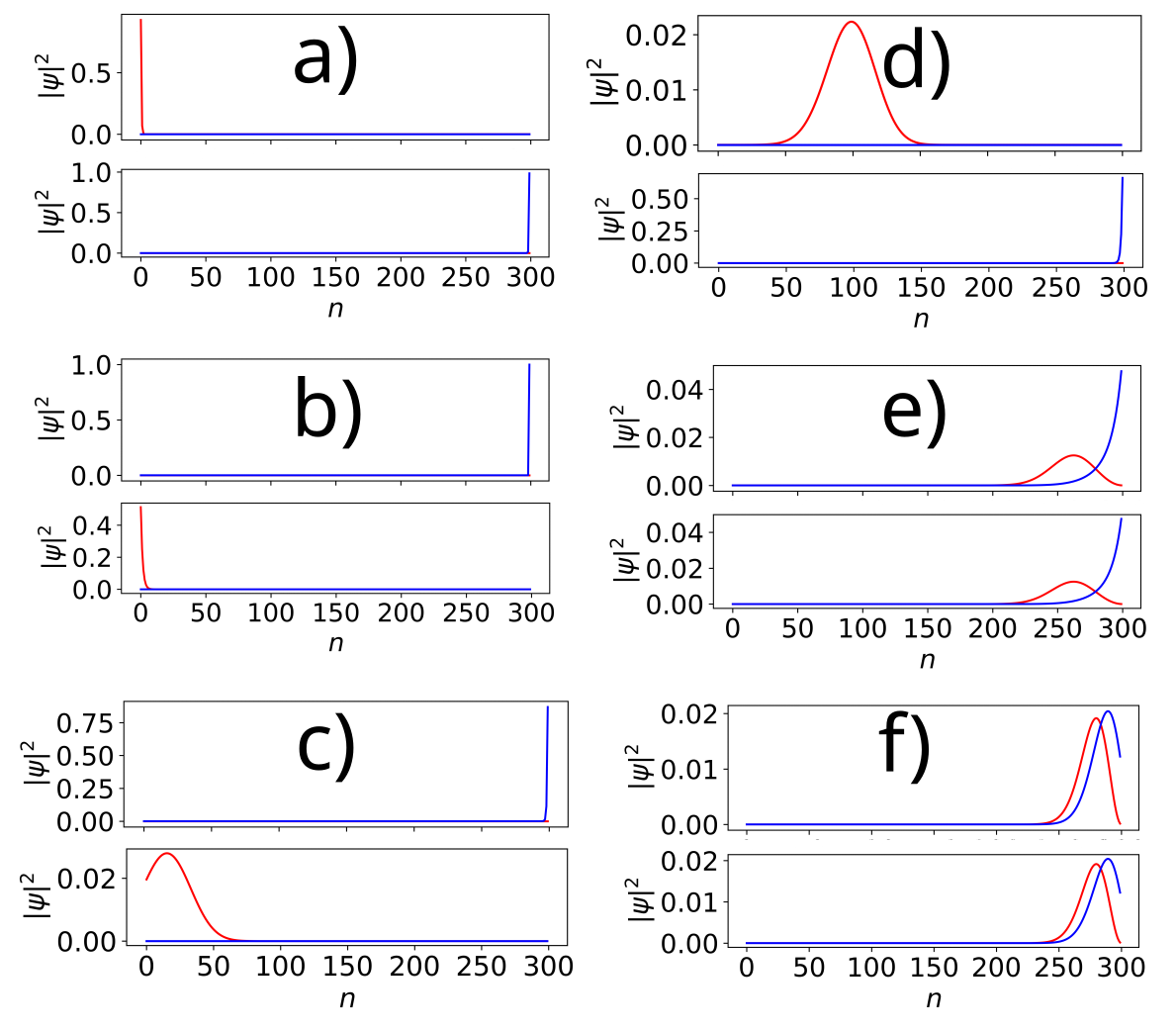}
\par\end{centering}
\caption{Evolution of the doubly degenerate edge mode into a pair of non-degenerate
Landau level edge state as $\delta$ is varied for parameters $t_{1}=t_{2}=\Delta_{1}=1$
and $\Delta_{2}=0.5$ in a chain cAA chain with 600 atoms, and a periodicity
or field $b=1/2001$. The phason is given by (a) $\delta/2\pi=0.59$,
(b) $\delta/2\pi=0.67$, (c) $\delta/2\pi=0.75$, (d) $\delta=0.80$,
(e) $\delta=0.88$, (f) $\delta=0.92$. The chiral symmetry of the
1D model initially ensures that the doubly degenerate edge modes at
$\varepsilon=0$ in each of the two distinct sublattices (represented
distinctly as blue and red curves) are located at both edges. By the
time the degeneracy is lifted by shifting the phason $\delta$, the
wave-function is still supported in both sublattices but with distinct
envelopes. One of the edge-states is progressively brought towards
the bulk and eventually both become localized at a single edge. \label{fig:Landau_Levels}}
\end{figure}

Finally, it is worth mentioning that that intuition about this model
can be gained by drawing an analogy to the well understood model of
the Half-integer quantum hall effect (HIQHE) in graphene. For instance,
if we pick $1/b=1/2001$, by computing the wave-function of edge states
for certain values of $\delta$, we see how, much like in the case
of zig-zag graphene ribbons, a zero-energy edge mode can turn into
a Landau-level edge state given the presence of Dirac cones in the
2D model. Additionally, the square root dispersion of the Landau levels,
characteristic to the HIQHE, also becomes clearly observable.  The dispersion in the presence of finite and small $b$ is showcased in Fig. \ref{fig:Landau_Levels},
together with the useful winding number invariant, which is unity
while the model hosts zero energy modes, and becomes zero after the
degeneracy is lifted by shifting the modulation.  The possibility of changing the edge localization of the wavefunction as $\delta$ is shifted is also illustrated in Fig. \ref{fig:Landau_levels}.

The difference relative to the 2D case of graphene, is, of course, the
fact that $\delta$ is a fixed parameter for a physical realization
of the cAA model, and hence the existence of edge modes of the 1D
model are also fixed, up to physical changes in $\delta$.

\begin{figure}
\begin{centering}
\includegraphics[width=0.45\textwidth]{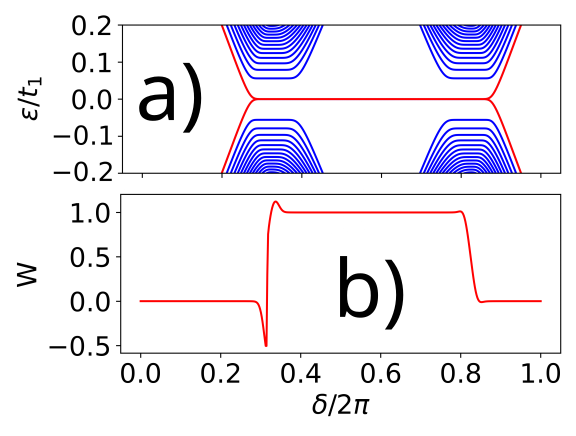}
\par\end{centering}
\caption{(a) Emergence of Landau levels in the cAA model in a topological semimetal
phase with two Dirac cones. We use the parameters $t_{1}=t_{2}=\Delta_{1}=1$
and $\Delta_{2}=0.5$ in a chain cAA chain with 600 atoms, and a periodicity
or field $b=1/2001$. (b) Winding number invariant, matching the presence
of doubly degenerate edge modes, and changing with the presence of
Dirac cones.\label{fig:finite_b_Winding}}
\end{figure}

\section{Conclusions and outlook}

In this work, we have introduced and studied the properties of a chiral
symmetric SSH-type model perturbed by an off-diagonal Aubry-André
(odAA) modulation, resulting in what we call the cAA model. The main
result of the work is the explicit demonstration that topological
invariants characteristic of different dimensionalities can coexist
in a physical way in a tight-binding purely one dimensional chain.
We believe that this result paves the way for a new series of topologically
non-trivial models with admixtures of different topological invariants,
thus increasing the tuneability and application range of topological
systems. Our idea is that by applying an incommensurate potential
to a system which already hosts some sort of topological invariant
due to the presence of a symmetry, the topology of the model can be
enhanced via the generation of synthetic dimensions with topologically
non-trivial degrees of freedom. This admixture of topological invariants
will then be manifest in the original system by the possibility of
introducing edge states of different character, such as the zero energy
edge modes, and Landau levels, as observed here in the cAA model.

To showcase this, we first analytically explored a limit of large
modulation periodicity of the cAA model, corresponding to what we
call the ``vanishing field limit''. Here, we have found many interesting
properties, such as the existence of a quantized winding number and
the presence of Dirac cones when a 2D crystalline superspace model
is constructed. In addition, we have provided a way to compute the
number of massless Dirac cones solely from the winding number, counting
the amount of times its value is changed as the AA modulation is shifted
relative to the lattice. We have found that $0,2$ or $4$ cones appear
in the vanishing field limit, located at variable positions along
certain lines in the superspace Brillouin zone. This
position depends on the relative strengths of hoppings and modulations. 

Then, we extended this methodology to the case of smaller odAA modulation
periodicity, by making use of topological markers. For periodicities
commensurate with the lattice and $1/b=p/q$, we have found the number
of massless Dirac cones in the superspace description is $0,2q$ or
$4q$, for small enough $q$. This value then saturates for $p/q$
approaching irrational values when the size of the unit cell induced
by the modulation becomes comparable with the physical system size. 

We have found that such a model can provide a platform for studying
two different kinds of integer quantum hall effect, namely when Dirac
cones are present (a HIQHE similar to the case of graphene) or absent
(the standard IQHE similar to the square lattice). We find that it
is possible to transition between the two by merging and splitting
Dirac cones when the system's parameters are varied. Both types of
IQHE are characterized by a Chern number, and when projected down
to physical space from the superspace model, can lead to the localization
of states at the edges, in correspondence with Landau levels in the
superspace model. 

The coexistence of zero-energy edge modes protected by the chiral
symmetry and Landau level edge states points to the exotic nature
of the cAA, and the explicit computation of topological invariants
all but confirms the coexistence, as we find that both invariants
$C$ and $W$ are necessary to characterize the potential existence
of edge states for a given physical realization of $\delta$ and Fermi
energy $E_{F}$. Furthermore, from the perspective of the 1D physical
system, this coexistence means that it is possible, for small periodicities,
to smoothly transform from 1D-type zero-energy edge modes located
at both edges, to 2D-type Landau-level like finite energy edge-states,
located at the same physical edge of the 1D chain.

Finally, much like other previously studied topological quasi-crystals,
an experimentally feasible implementation of the cAA could be studied
using photonic crystals, where the hopping modulation could be achieved
by controlling the spacing between distinct fiber optics \citep{Kraus_2012,Longhi2020}.
In this setting, minimal changes to the way waveguides are spaced
are necessary to realize the cAA model. Besides this one, many other
alternatives exist, such as: ultra-cold atom platforms, where generalized
AA type models have been implemented in the past \citep{Kunal_2019,Li_2022};
quantum simulators, using, for instance superconducting qubits \citep{Li_2023},
or other types of quantum computational devices; and magnetic systems
\citep{Lado2019}, with proposals existing for the realization of
Aubry-André type systems taking advantage of twisted van der Waals
heterostructures. 

\section*{Acknowledgments}

T.V.C.A. acknowledges support by the Portuguese Foundation for Science
and Technology (FCT) in the framework of the project CERN/FIS-COM/0004/2021
and the hospitality of LIP where this work was conducted. T.V.C.A.
also acknowledges the the computational resources provided by the
Aalto Science-IT project. N.M.R.P. acknowledges support by the Portuguese
Foundation for Science and Technology (FCT) in the framework of the
Strategic Funding UIDB/04650/2020, COMPETE 2020, PORTUGAL 2020, FEDER,
and through projects PTDC/FIS-MAC/2045/2021, EXPL/FIS-MAC/0953/ 2021,
and from the European Commission through the project Graphene Driven
Revolutions in ICT and Beyond (Ref. No. 881603, CORE 3). N.M.R.P.
also acknowledges the Independent Research Fund Denmark (grant no.
2032-00045B) and the Danish National Research Foundation (Project No.\textasciitilde DNRF165).

\appendix

\section{Details of the superspace model in the vanishing field limit}

In the main text we have presented the superspace model in the vanishing
field limit without much discussion of its derivation. In this Appendix,
we provide a quick and simple derivation, starting from the Hamiltonian
which realizes, for a fixed $\delta$, a physical implementation of
the cAA model. It reads

\begin{align}
H(\delta)= & -\sum_{n}\left[t_{1}+\Delta_{1}\cos\left(2\pi bn-\delta\right)\right]a_{n}^{\dagger}b_{n}\nonumber \\
 & +\left[t_{2}+\Delta_{2}\cos\left(2\pi bn-\delta\right)\right]b_{n}^{\dagger}a_{n+1}+\text{h.c.},
\end{align}
The very simple trick which allows us to generate the 2D superspace
model is to expand the cosine modulations in terms of exponentials
and allow, in the vanishing field limit $b\to0$, the dependence on
$b$ to drop out from the Hamiltonian, yielding a $b-$independent
Hamiltonian. 

\begin{align}
H(\delta)= & -\sum_{n}\left[t_{1}+\frac{\Delta_{1}}{2}\left(e^{i\delta}+e^{-i\delta}\right)\right]a_{n}^{\dagger}b_{n}\nonumber \\
 & +\left[t_{2}+\frac{\Delta_{2}}{2}\left(e^{i\delta}+e^{-i\delta}\right)\right]b_{n}^{\dagger}a_{n+1}+\text{h.c.},
\end{align}
The fact that this form of the Hamiltonian is a good description of
the model can be checked numerically, by setting, for instance $b=1/1001$,
which produces results for the 1D topology practically indistinct
from the vanishing field limit. The idea now is to imagine that $\delta$
plays the role of a momentum variable, and promote the creation and
annihilation operators into objects with two indices. The first is
$n$, the lattice site along physical space which for the sake of
clarity we rename to $i$ to better signify this change to a 2D model,
and the second one is $j$ which is the position along the synthetic
dimension, conjugate to the momentum $\delta$. So, our aim is to
transform $a_{n}^{\dagger}$ and $b_{n}^{\dagger}$ as

\begin{align}
a_{n}^{\dagger} & \to\sum_{j}a_{i,j}^{\dagger}e^{-i\delta j},\\
b_{n}^{\dagger} & \to\sum_{j}b_{i,j}^{\dagger}e^{-i\delta j},
\end{align}
and account for the already existing exponentials $e^{\pm i\delta}$
as indicators of the existence of couplings in the synthetic direction.
To see why this can be done, simply note that it is possible to write

\begin{align}
a_{i}^{\dagger}b_{i}e^{-i\delta} & =\sum_{j,j'}b_{i,j'}^{\dagger}e^{-i\delta j'}a_{i,j}e^{i\delta j}e^{-i\delta}.\nonumber \\
 & =\sum_{j,j'}b_{i,j'}^{\dagger}a_{i,j}e^{-i\delta\left(j'+1\right)}e^{i\delta j}\nonumber \\
 & =\sum_{j}b_{i,j}^{\dagger}a_{i,j+1}.\label{eq:Couplings}
\end{align}
Thus, the Hamiltonian of the system can be brought into 2D by splitting
up the terms proportional to $e^{\pm i\delta}$, adding a summation
in the synthetic dimension, an adjusting for the synthetic couplings
generated as in equation (\ref{eq:Couplings}). This procedure results
in equation (\ref{eq:2Dreal_space}) of the main text, which we repeat
here for convenience.

\begin{align}
H_{\text{2D}}= & -\sum_{i,j}t_{1}a_{i,j}^{\dagger}b_{i,j}+\Delta_{1}'a_{i,j}^{\dagger}b_{i,j+1}+\text{h.c.}\nonumber \\
 & -\sum_{i,j}t_{1}b_{i,j}^{\dagger}a_{i+1,j}+\Delta_{2}'b_{i,j}^{\dagger}a_{i+1,j+1}+\text{h.c.},
\end{align}
This concludes our small derivation of the 2D superspace model in
a ``real space'' representation. The generated hoppings are ``diagonal''
in the 2D lattice, as illustrated in Fig. \ref{fig:Illustration}
of the main text. Instead of choosing this representation, we can
instead go into momentum space in both physical and synthetic dimensions.
This amounts to performing the transformation

\begin{align}
a_{n}^{\dagger} & \to\sum_{k}a_{k}^{\dagger}e^{ikn},\\
b_{n}^{\dagger} & \to\sum_{k}b_{k}^{\dagger}e^{ikn},
\end{align}
which leads to, summing over all possible values of $\delta$ in order
to take into account the synthetic dimension
\begin{align}
H_{2D} & =\sum_{k,\delta}\left[t_{1}+\Delta_{1}\cos\delta\right]a_{k}^{\dagger}b_{k}\nonumber \\
 & +\left[t_{2}+\Delta_{2}\cos\delta\right]b_{k}^{\dagger}a_{k}e^{ik}+\text{h.c.},
\end{align}
A more convenient form is the Hamiltonian matrix which was described
in the main text

\begin{align}
H_{2D}(\boldsymbol{k})= & -\left[t_{1}+\Delta_{1}'\cos\delta+\left(t_{2}+\Delta_{2}'\cos\delta\right)\cos k\right]\sigma_{x}\nonumber \\
 & -\left[\left(t_{2}+\Delta_{2}'\cos\delta\right)\sin k\right]\sigma_{y}.
\end{align}
 It is also worth remarking that the role played by the additional
exponentials $e^{\pm i2\pi bn}$ is to introduce additional phases
along these hoppings, which explicitly break translational symmetry.
Despite this, the position dependence on $n$ is precisely what makes
the system exactly mappeable to an IQHE model in the Landau gauge.
As we show in the main text, this limit can be analyzed numerically
with topological markers.

\section{Dimensional reduction and an analogy with graphene}

Underlying the cAA model and the richness of topological quasicrystals
in general is the idea of synthetic dimensions. The procedure of mapping
the cAA model into a superspace model involves taking a model parameter,
in our case the modulation shift, and interpreting it as a momentum
variable, thus generating a higher dimensional model with an additional
synthetic dimension. The inverse procedure can be performed, and the
role of this Appendix is to illustrate it via the example of graphene.
We start with a physical 2D model and perform a process of dimensional
reduction, by reinterpreting a momentum variable as an adjustable
parameter. 

\begin{figure}
\begin{centering}
\includegraphics[scale=0.4]{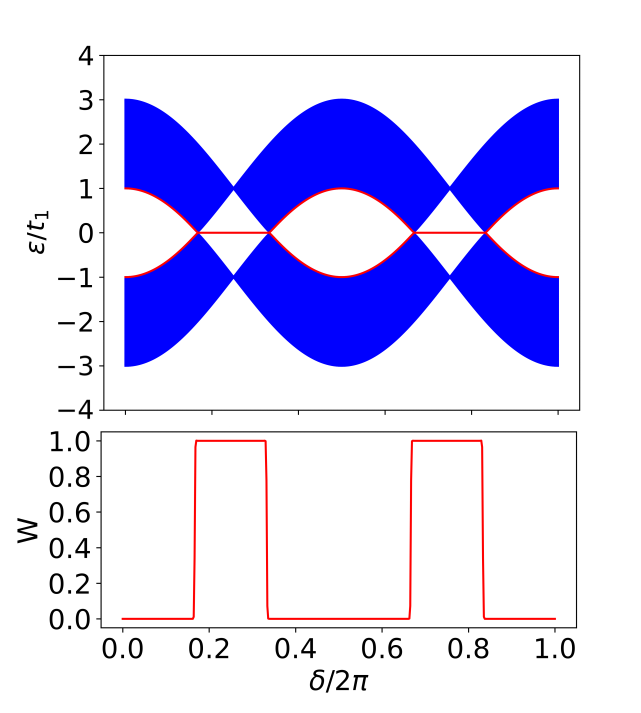}
\par\end{centering}
\caption{(a) Energy spectrum for a graphene Zig-zag ribbon, simulated using
the cAA model. (b) Winding number of the SSH model resulting from
dimensional reduction of the zig-zag graphene ribbon. We have $W(\delta)=1$
when doubly degenerate edge states are present, and $W(\delta)=0$
when the  states wander into the bulk. \label{fig:(a)-Energy-spectrum}}
\end{figure}

A simple example of this procedure, but which is often overlooked
is a zig-zag edge graphene ribbon. Like the cAA model, zig-zag graphene
has two sublattices $A$ and $B$ and, for a ribbon with $N$ unit
cells in the finite direction, the Hamiltonian is of form

\begin{equation}
H_{\text{g}}=-t\sum_{i=1}^{\infty}\sum_{j=1}^{N}a_{i,j}^{\dagger}b_{i,j}+b_{i,j}^{\dagger}a_{i,j+1}+b_{i,j}^{\dagger}a_{i+1,j}+\text{h.c.}.
\end{equation}
It is then possible to Fourier transform this Hamiltonian along the
infinite direction, introducing the momentum variable $k$, resulting
in

\begin{equation}
H_{g}=-t\sum_{k}\sum_{j=1}^{N}2\cos\left(\frac{\sqrt{3}}{2}k\right)a_{k,j}^{\dagger}b_{k,j}+b_{k,j}^{\dagger}a_{k,j+1}+\text{h.c.}.
\end{equation}
If we ignore the summation in $k$, the Hamiltonian is simply that
of a finite SSH chain with hoppings modulated by the momentum $k$.
Thus, from the 2D graphene lattice we have constructed a modulated
1D SSH model via a process of dimensional reduction. A winding number
can be associated to this SSH model, and the winding number indicates
the presence of the characteristic degenerate edge modes at zero energy.
This is a topological interpretation of this phenomena which is often
overlooked. In Fig. \ref{fig:(a)-Energy-spectrum}, we present the
result of using the winding number topological marker presented in
the main text to compute the topological invariant of the SSH model
resulting from the dimensional reduction procedure when applied to
a graphene zig-zag ribbon.

Finally, note that graphene can actually be mapped to a particular
parameter set of the cAA model. By picking physical hoppings $t_{1}=\Delta_{2}=0$,
$t_{2}=t$, and $\Delta_{1}=2t$, as well as a value of $b=0$ and
identifying $\sqrt{3}k/2$ with the shift $\delta$, we can simulate
graphene in the cAA model. Since graphene appears in the limit of
$b=0$, the parameter $\delta$ does not exactly play the role of
a shift of the potential, but rather, by itself, it globally reduces
or increases the strength of the intra-cell hopping. The connection
between these models therefore definitely exists, however, the fact
that no actual Aubry-André periodic modulation is present removes
many of the puzzling features of the cAA model. Nonetheless, this
correspondence establishes that the concept of synthetic dimensions
can be utilized, in principle, to simulate a graphene lattice in a
1D model, by accessing slices of its spectrum, and tuning the tight-binding
hopping strengths in order to move around the momentum direction. 

\begin{thebibliography}{0}%
\makeatletter
\providecommand \@ifxundefined [1]{%
 \@ifx{#1\undefined}
}%
\providecommand \@ifnum [1]{%
 \ifnum #1\expandafter \@firstoftwo
 \else \expandafter \@secondoftwo
 \fi
}%
\providecommand \@ifx [1]{%
 \ifx #1\expandafter \@firstoftwo
 \else \expandafter \@secondoftwo
 \fi
}%
\providecommand \natexlab [1]{#1}%
\providecommand \enquote  [1]{``#1''}%
\providecommand \bibnamefont  [1]{#1}%
\providecommand \bibfnamefont [1]{#1}%
\providecommand \citenamefont [1]{#1}%
\providecommand \href@noop [0]{\@secondoftwo}%
\providecommand \href [0]{\begingroup \@sanitize@url \@href}%
\providecommand \@href[1]{\@@startlink{#1}\@@href}%
\providecommand \@@href[1]{\endgroup#1\@@endlink}%
\providecommand \@sanitize@url [0]{\catcode `\\12\catcode `\$12\catcode
  `\&12\catcode `\#12\catcode `\^12\catcode `\_12\catcode `\%12\relax}%
\providecommand \@@startlink[1]{}%
\providecommand \@@endlink[0]{}%
\providecommand \url  [0]{\begingroup\@sanitize@url \@url }%
\providecommand \@url [1]{\endgroup\@href {#1}{\urlprefix }}%
\providecommand \urlprefix  [0]{URL }%
\providecommand \Eprint [0]{\href }%
\providecommand \doibase [0]{https://doi.org/}%
\providecommand \selectlanguage [0]{\@gobble}%
\providecommand \bibinfo  [0]{\@secondoftwo}%
\providecommand \bibfield  [0]{\@secondoftwo}%
\providecommand \translation [1]{[#1]}%
\providecommand \BibitemOpen [0]{}%
\providecommand \bibitemStop [0]{}%
\providecommand \bibitemNoStop [0]{.\EOS\space}%
\providecommand \EOS [0]{\spacefactor3000\relax}%
\providecommand \BibitemShut  [1]{\csname bibitem#1\endcsname}%
\let\auto@bib@innerbib\@empty
\end{thebibliography}%


\begin{thebibliography}{10}

\bibitem{Shechtman_2013}
Dan Shechtman, I.~Blech, D.~Gratias, and John~W. Cahn.
\newblock {\em Metallic Phase with Long-Range Orientational Order and No
  Translational Symmetry}.
\newblock 2013.

\bibitem{Macia_2006}
Enrique Maci\'{a}.
\newblock The role of aperiodic order in science and technology.
\newblock {\em Reports on Progress in Physics}, 2006.

\bibitem{Steinhardt_1987}
Paul~J. Steinhardt and Stellan Ostlund.
\newblock {\em The physics of quasicrystals}.
\newblock 1987.

\bibitem{Senechal_1995}
Marjorie Senechal.
\newblock {\em Quasicrystals and geometry}.
\newblock 1995.

\bibitem{Duneau_null}
Michel Duneau and Andr\'{e} Katz.
\newblock Quasiperiodic patterns.
\newblock {\em Physical Review Letters}, 1985.

\bibitem{Elser_1985}
Veit Elser and Christopher~L. Henley.
\newblock Crystal and quasicrystal structures in al-mn-si alloys.
\newblock {\em Physical Review Letters}, 1985.

\bibitem{Jagannathan_2021}
Anuradha Jagannathan.
\newblock The fibonacci quasicrystal: Case study of hidden dimensions and
  multifractality.
\newblock {\em Reviews of Modern Physics}, 2021.

\bibitem{Bruijn_1981}
N.~G. de~Bruijn.
\newblock {\em Algebraic theory of Penrose's non-periodic tilings of the plane.
  II}.
\newblock 1981.

\bibitem{Moon_2019}
Pilkyung Moon, Mikito Koshino, and Young-Woo Son.
\newblock Quasicrystalline electronic states in ${30}^{\ensuremath{\circ}}$
  rotated twisted bilayer graphene.
\newblock {\em Phys. Rev. B}, 99:165430, Apr 2019.

\bibitem{Amorim2023}
Bruno Amorim, Flavio Riche, Eduardo Castro, Pedro Ribeiro, and Miguel
  Gon{\c{c}}alves.
\newblock Incommensurability enabled quasi-fractal order in 1d narrow-band
  moir{\'{e}} systems.
\newblock June 2023.

\bibitem{Crosse2021}
J.~A. Crosse and Pilkyung Moon.
\newblock Quasicrystalline electronic states in twisted bilayers and the
  effects of interlayer and sublattice symmetries.
\newblock {\em Physical Review B}, 103(4), January 2021.

\bibitem{Wu_2021}
Ang-Kun Wu.
\newblock Fractal spectrum of the aubry-andre model.
\newblock {\em arXiv:2109.07062v2}, 2021.

\bibitem{Dominguez-Castro_2018}
G.A. Dom\'{i}nguez-Castro and R.~Paredes.
\newblock The aubry-andr\'{e} model as the hobbyhorse for understanding
  localization phenomenon.
\newblock {\em arXiv: Quantum Gases}, 2018.

\bibitem{Kraus_2012}
Yaacov~E. Kraus, Yoav Lahini, Zohar Ringel, Mor Verbin, and Oded Zilberberg.
\newblock Topological states and adiabatic pumping in quasicrystals.
\newblock {\em Bulletin of the American Physical Society}, 2012.

\bibitem{Kraus_2013}
Yaacov~E. Kraus, Zohar Ringel, and Oded Zilberberg.
\newblock Four-dimensional quantum hall effect in a two-dimensional
  quasicrystal.
\newblock {\em Physical Review Letters}, 2013.

\bibitem{Lohse_2018}
Michael Lohse, Christian Schweizer, Hannah~M. Price, Oded Zilberberg, and
  Immanuel Bloch.
\newblock Exploring 4d quantum hall physics with a 2d topological charge pump.
\newblock {\em Nature}, 2018.

\bibitem{Prodan_2015}
Emil Prodan.
\newblock Virtual topological insulators with real quantized physics.
\newblock {\em Physical Review B}, 2015.

\bibitem{Hofstadter_1976}
Douglas~R. Hofstadter.
\newblock Energy levels and wave functions of bloch electrons in rational and
  irrational magnetic fields.
\newblock {\em Physical Review B}, 1976.

\bibitem{Klitzing_1980}
Klaus von Klitzing, G.~Dorda, and Michael Pepper.
\newblock New method for high-accuracy determination of the fine-structure
  constant based on quantized hall resistance.
\newblock {\em Physical Review Letters}, 1980.

\bibitem{Hatsugai_1993}
Yasuhiro Hatsugai.
\newblock Chern number and edge states in the integer quantum hall effect.
\newblock {\em Physical Review Letters}, 1993.

\bibitem{Thouless_82}
D.~J. Thouless, M.~Kohmoto, M.~P. Nightingale, and M.~den Nijs.
\newblock Quantized hall conductance in a two-dimensional periodic potential.
\newblock {\em Phys. Rev. Lett.}, 49:405--408, Aug 1982.

\bibitem{Else_2021}
Dominic~V. Else, Sheng-Jie Huang, Abhinav Prem, and Andrey Gromov.
\newblock Quantum many-body topology of quasicrystals.
\newblock {\em Physical Review X}, 2021.

\bibitem{Zurita2021}
Juan Zurita, Charles Creffield, and Gloria Platero.
\newblock Tunable zero modes and quantum interferences in flat-band topological
  insulators.
\newblock {\em Quantum}, 5:591, November 2021.

\bibitem{Ganeshan2015}
Sriram Ganeshan and S.~Das Sarma.
\newblock Constructing a weyl semimetal by stacking one-dimensional topological
  phases.
\newblock {\em Physical Review B}, 91(12), March 2015.

\bibitem{Rosenberg2022}
Peter Rosenberg and Efstratios Manousakis.
\newblock Topological superconductivity in a two-dimensional weyl {SSH} model.
\newblock {\em Physical Review B}, 106(5), August 2022.

\bibitem{Longhi2020}
Stefano Longhi.
\newblock Topological Anderson phase in quasi-periodic waveguide lattices.
\newblock {\em Optics Letters}, 45(14), July 2020.

\bibitem{Su_1979}
Wu-Pei Su, J.~R. Schrieffer, and Alan~J. Heeger.
\newblock Solitons in polyacetylene.
\newblock {\em Physical Review Letters}, 1979.

\bibitem{Lado2019}
J.~L. Lado and Oded Zilberberg.
\newblock Topological spin excitations in harper-heisenberg spin chains.
\newblock {\em Physical Review Research}, 1(3), October 2019.

\bibitem{Kraus_2012_1}
Yaacov~E. Kraus and Oded Zilberberg.
\newblock Topological equivalence between the fibonacci quasicrystal and the
  harper model.
\newblock {\em Physical Review Letters}, 2012.

\bibitem{Neto_2009}
A.~H.~Castro Neto, Francisco Guinea, N.~M.~R. Peres, Kostya~S. Novoselov, and
  Andre~K. Geim.
\newblock The electronic properties of graphene.
\newblock {\em Reviews of Modern Physics}, 2009.

\bibitem{KITAEV_2006_2}
Alexei Kitaev.
\newblock Anyons in an exactly solved model and beyond.
\newblock {\em Annals of Physics}, 321(1):2--111, 2006.
\newblock January Special Issue.

\bibitem{Bianco_Resta_2011}
Raffaello Bianco and Raffaele Resta.
\newblock Mapping topological order in coordinate space.
\newblock {\em Physical Review B}, 84:241106, Dec 2011.

\bibitem{Gersdorff_2021}
Gero von Gersdorff, Shahram Panahiyan, and Wei Chen.
\newblock Unification of topological invariants in dirac models.
\newblock {\em arXiv: Mesoscale and Nanoscale Physics}, 2021.

\bibitem{Chen_2022}
Wei Chen.
\newblock Universal topological marker.
\newblock {\em Physical review B}, 2022.

\bibitem{Sykes_2022}
Joseph Sykes and Ryan Barnett.
\newblock 1d quasicrystals and topological markers.
\newblock {\em Materials for quantum technology}, 2022.

\bibitem{Altland_Zirnbauer_1997}
Alexander Altland and Martin~R. Zirnbauer.
\newblock Nonstandard symmetry classes in mesoscopic normal-superconducting
  hybrid structures.
\newblock {\em Phys. Rev. B}, 55:1142--1161, Jan 1997.

\bibitem{Ryu_2010}
Shinsei Ryu, Andreas~P Schnyder, Akira Furusaki, and Andreas W~W Ludwig.
\newblock Topological insulators and superconductors: tenfold way and
  dimensional hierarchy.
\newblock {\em New Journal of Physics}, 12(6):065010, jun 2010.

\bibitem{Nielsen1981}
H.~B. Nielsen and M.~Ninomiya.
\newblock Absence of neutrinos on a lattice.
\newblock {\em Nuclear Physics B}, 185(1):20--40, July 1981.

\bibitem{Nielsen1981_2}
H.B. Nielsen and M.~Ninomiya.
\newblock Absence of neutrinos on a lattice.
\newblock {\em Nuclear Physics B}, 193(1):173--194, December 1981.

\bibitem{Krivosenko2021}
Yury~S Krivosenko, Ivan~V Iorsh, and Ivan~A Shelykh.
\newblock Bosonic hofstadter butterflies in synthetic antiferromagnetic
  patterns.
\newblock {\em Journal of Physics: Condensed Matter}, 33(13):135802, January
  2021.

\bibitem{Thouless_1983}
David~J. Thouless.
\newblock Quantization of particle transport.
\newblock {\em Physical Review B}, 1983.

\bibitem{Vanderbilt_2018}
D.~Vanderbilt.
\newblock {\em Berry Phases in Electronic Structure Theory}.
\newblock 2018.

\bibitem{Spaldin_2012}
Nicola~A. Spaldin.
\newblock A beginner's guide to the modern theory of polarization.
\newblock {\em Journal of Solid State Chemistry}, 2012.

\bibitem{Kunal_2019}
Kunal~K. Das and Jacob Christ.
\newblock Realizing the harper model with ultracold atoms in a ring lattice.
\newblock {\em Phys. Rev. A}, 99:013604, Jan 2019.

\bibitem{Li_2022}
Yi~Li, Jia-Hui Zhang, Feng Mei, Jie Ma, Liantuan Xiao, and Suotang Jia.
\newblock Generalized aubry-andr\'{e}-harper models in optical superlattices.
\newblock {\em Chinese Physics Letters}, 39(6):063701, jun 2022.

\bibitem{Li_2023}
Hao Li, Yong-Yi Wang, Yun-Hao Shi, Kaixuan Huang, Xiaohui Song, Gui-Han Liang,
  Zheng-Yang Mei, Bozhen Zhou, He~Zhang, Jia-Chi Zhang, Shu Chen, S.~P. Zhao,
  Ye~Tian, Zhan-Ying Yang, Zhongcheng Xiang, Kai Xu, Dongning Zheng, and Heng
  Fan.
\newblock Observation of critical phase transition in a generalized
  aubry-andr{\'{e}}-harper model with superconducting circuits.
\newblock {\em npj Quantum Information}, 9(1), April 2023.

\end{thebibliography}
\end{document}